\documentclass[aps]{revtex4}
\usepackage{amsmath}
\usepackage{amssymb,epsf}

\begin{document}

\title{Thermodynamic description and quasinormal modes of adS black holes\\
in Born-Infeld massive gravity with a non-abelian hair}
\author{Seyed Hossein Hendi$^{1,2}$\footnote{%
email address: hendi@shirazu.ac.ir} and Mehrab Momennia$^{1}$\footnote{%
email address: m.momennia@shirazu.ac.ir}} \affiliation{$^1$
Physics Department and Biruni Observatory, College of Sciences,
Shiraz
University, Shiraz 71454, Iran\\
$^2$ Research Institute for Astronomy and Astrophysics of Maragha (RIAAM),
P.O. Box 55134-441, Maragha, Iran}

\begin{abstract}
We construct a new class of asymptotically (a)dS black hole solutions of
Einstein-Yang-Mills massive gravity in the presence of Born-Infeld nonlinear
electrodynamics. The obtained solutions possess a Coulomb electric charge,
massive term and a non-abelian hair as well. We calculate the conserved and
thermodynamic quantities, and investigate the validity of the first law of
thermodynamics. Also, we investigate thermal stability conditions by using
the sign of heat capacity through canonical ensemble. Next, we consider the
cosmological constant as a thermodynamical pressure and study the van der
Waals like phase transition of black holes in the extended phase space
thermodynamics. Our results indicate the existence of a phase transition
which is affected by the parameters of theory. Finally, we consider a
massless scalar perturbation in the background of asymptotically adS
solutions and calculate the quasinormal modes by employing the
pseudospectral method. The imaginary part of quasinormal frequencies is the
time scale of a thermal state (in the conformal field theory) for the
approach to thermal equilibrium.
\end{abstract}

\maketitle


\section{Introduction}

General relativity (GR) of Einstein is one of the most successful theories
in theoretical physics. It gave a more insightful picture to understanding
the gravity and solved some unanswered problems. Despite its amazing
achievements to justify some phenomena, such as perihelion precession of
Mercury, deflection of light, and gravitational redshift, there are still
some unsolved problems in the universe. Among them, one can point out the
hierarchy problem, the cosmological constant problem, and the late time
accelerated expansion of the Universe. This shows that GR is not the final
theory and it is logical to search for a more general and complete theory
which be able to solve unanswered problems. GR is a theory which describes
massless spin-$2$ particles \cite{massless}. In order to generalize GR into
a more effective theory, one can give mass to massless spin-$2$ particles
and consider them as massive spin-$2$ particles. Such a theory is called a
massive theory of gravity.

One of the most well known theories of massive gravity is called dRGT model
and has been introduced by de Rham, Gabadadze, and Tolley \cite{dRGT,dRGTPRL}
which added a potential contribution to the Einstein-Hilbert action. This
potential gives graviton a mass and modifies the dynamics of GR in the IR
limit. The authors indicated that the theory is ghost free in the decoupling
limit to all orders of nonlinearities. On the other hand, massive couplings $%
c_{i}$'s are arbitrary constants and by choosing different massive
couplings, different theories can be obtained. Hassan and Rosen improved the
previous result to all orders in $4$-dimension \cite{HassanPRL}. They
confirmed that any pathological Boulware--Deser ghost is eliminated at the
full nonlinear level due to the Hamiltonian constraint and generalize their
ghost analysis to the most general case for arbitrary massive couplings $%
c_{i}$'s. It has been also shown that the massive gravity with a general
reference metric is ghost free \cite{Hassan}. The dRGT massive gravity is
almost a successful model in a sense that it does not lead to van
Dam-Veltman-Zakharov discontinuity, it is free of Boulware--Deser ghost, and
it can be used in higher dimensions with admissible validity. Nevertheless,
the cosmological solutions do not admit flat FLRW metric and theory exhibits
a discontinuity at the flat FLRW limit \cite{Amico,Gumrukcuoglu} or the
model meets instabilities \cite{GumrukcuogluPRD,AmicoCQG,Chullaphan}.

On the other hand, the dRGT model has different modifications which are
based on the definition of the reference metric. The most successful one has
been introduced by Vegh \cite{Vegh} with the motivation of breaking the
translational symmetry. In other words, this model provides an effective
bulk description in which momentum is not conserved anymore, and therefore,
it includes holographic momentum dissipation. This property is what people
needed to study physical systems in the context of gauge/gravity duality. In
addition, it was shown that this model is ghost free and stable \cite{Zhang}%
. The static black hole solutions and magnetic solutions in the presence of
this model of massive gravity have been investigated in \cite%
{CaiMassive,HendiMassive,GhoshMassive,HendiMassiveBTZ} and \cite%
{MagneticMassive,ThreeMagneticMassive}, respectively. Moreover, the
thermodynamic properties and van der Waals like phase transition of black
holes have been studied \cite%
{HendiMassive,Xu,HendiBerlin,HendiMann,HendiPLB2017}. From the cosmological
point of view, it has been shown that it is possible to remove the big bang
singularity \cite{Nonsingular}. In addition, the behavior of different
holographic quantities has been investigated in \cite%
{Vegh,RADavison,MBlake,LAlberte,XXZeng,Dehyadegari}.

On the other hand, the existence of some limitations in the Maxwell theory
motivates one to consider nonlinear electrodynamics (NED) \cite%
{Heisenberg,Yajima,Schwinger,Lorenci,LorenciKlippert,NovelloBittencourt,Novello,DelphenichQED,Delphenich}%
. Moreover, it was shown that NED can remove both the big bang and black
hole singularities \cite%
{BeatoGRG,BeatoGarcia,LorenciNovello,Dymnikova,Corda,CordaMosquera}. In
addition, the effects of NED are important in superstrongly magnetized
compact objects \cite{Mosquera,MosqueraSalim,Birula}. Considering GR coupled
to NED attracts attention due to its specific properties in gauge/gravity
coupling. Besides, NED theories are richer than the linear Maxwell theory
and in some special cases, they reduce to the Maxwell electrodynamics.

One of the most interesting NED theories has been introduced by Born and
Infeld \cite{Born,BornInfeld} in order to remove the divergency of self
energy of a point-like charge at the origin. The Lagrangian of Born-Infeld
(BI) nonlinear gauge field is given by%
\begin{equation}
\mathcal{L}_{BI}(\mathcal{F}_{M})=4\beta ^{2}\left( 1-\sqrt{1+\frac{\mathcal{%
F}_{M}}{2\beta ^{2}}}\right) ,  \label{BI}
\end{equation}%
where $\beta $\ is BI nonlinearity parameter, $\mathcal{F}_{M}=F_{\mu \nu
}F^{\mu \nu }$\ is the Maxwell invariant, $F_{\mu \nu }=2\nabla _{\lbrack
\mu }A_{\nu ]}$\ is the Faraday tensor, and $A_{\nu }$ is the gauge
potential. Using the expansion of this Lagrangian for a large value of
nonlinearity parameter leads to the Maxwell linear Lagrangian%
\begin{equation}
\mathcal{L}_{BI}(\mathcal{F}_{M})=-\mathcal{F}_{M}+\frac{\mathcal{F}_{M}^{2}%
}{8\beta ^{2}}+\mathcal{O}\left( \frac{1}{\beta ^{4}}\right) ,  \label{SerBI}
\end{equation}%
in which we receive the Maxwell Lagrangian at $\beta \rightarrow \infty $.
BI NED arises in the low-energy limit of the open string theory \cite%
{Fradkin,Matsaev,Bergshoeff,Callan,Andreev,Leigh}. From the AdS/CFT
correspondence point of view, it has been shown that, unlike gravitational
correction, higher derivative terms of nonlinear electrodynamics do not have
effect on the ratio of shear viscosity over entropy density \cite{ADSBI}.
Besides, NED theories make crucial effects on the condensation of the
superconductor and its energy gap \cite{superconductor,superconductorJing}.
GR in the presence of BI NED has been investigated for static black holes
\cite%
{Dehghani,DehghaniHendi,Allahverdizadeh,Zou,Mazharimousavi,Chemissany,Myung,Miskovic,Fernando,CaiPang,Cataldo,HendiENrainbow}%
, wormholes \cite{Lu,MHDehghaniHendi,Eiroa,Hendi}, rotating black objects
\cite%
{HHendi,DehghaniHendiJCAP,DehghaniHendiInt,DehghaniRastegar,HendiPRD,Ferrari}%
, and superconductors \cite{superconductorJing,Yao,Gangopadhyay,Jing}. In
addition, black hole solutions and their van der Waals like behavior in
massive gravity coupled to BI NED have been studied in \cite%
{HendiMassive,MingZhang}.

On the other hand, in addition to the Maxwell field, one can consider the
non-abelian Yang-Mills (YM) field as matter source coupled to gravity. The
presence of non-abelian gauge fields in the spectrum of some string models
motivates us to consider them coupled to GR. In addition, the YM equations
are present in the low energy limit of these models. Considering the YM
field coupled with gravity violates the black hole uniqueness theorem and
leads to hairy black holes. In such a situation, the field equations become
highly nonlinear so that the early attempts for finding the black hole
solutions in YM theory were performed numerically.

Nevertheless, Yasskin found the first analytic black hole solutions by using
Wu-Yang ansatz \cite{Yasskin}. Then, the black hole solutions in YM theory
have been generalized to Gauss--Bonnet and Lovelock gravity in \cite%
{MazharimousaviGB2007,MazharimousaviGB2008} and \cite%
{MazharimousaviLov,MazharimousaviLov2009}, respectively. In addition, black
holes have been investigated in non-abelian generalization of BI NED in
Einstein gravity \cite{Wirschins} and regular black holes have been obtained
in \cite{Beato,LemosZanchin,Bolokhov,Ma}. Furthermore, hairy black holes
coupled to YM field have been studied in \cite%
{Winstanley,Bjoraker,BjorakerHosotani}. Nonminimal Einstein-Yang-Mills (EYM)
solutions have been investigated for regular black holes \cite%
{BalakinLemos,BalakinLemosPRD}, wormholes \cite{BalakinSushkov,BalakinZayats}%
, and monopoles \cite{BalakinZayatsPLB,BalakinDehnen}. Thermodynamics and $%
P-V$ criticality of EYM black holes in gravity's rainbow have been explored
in \cite{YMrainbow}. Besides, the solutions of EYM-dilaton theory have been
considered in \cite{Lavrelashvili,Donets,Bizon,Torii,Brihaye,Radu,DehghaniYM}%
. In addition, black holes and their van der Waals like phase transition in
Gauss--Bonnet-massive gravity in the presence of YM field have been
investigated in \cite{Meng}.

The purpose of this paper is obtaining the exact black hole solutions of
Einstein-Massive theory in the presence of YM and BI NED fields\ (which is a
more general solution compared with Reissner-Nordstr\"{o}m,
Einstein-Born-Infeld \cite{EBI}, Einstein-Yang-Mills \cite{Yasskin},
Einstein-massive gravity \cite{CaiMassive}, Einstein-Born-Infeld-massive
gravity \cite{HendiMassive} and etc.), and also, studying the thermal
stability and phase transition of these black holes. Besides, we consider
the massless scalar perturbations in the background of asymptotically adS
solutions and calculate the quasinormal modes by employing the
pseudospectral method. We investigate the effects of the free parameters on
the quasinormal modes and dynamical stability. We also show that how the
free parameters affect the time scale that a thermal state in conformal
field theory (CFT) needs to pass to meet the thermal equilibrium.


\section{field equations and black hole solutions \label{FE}}

Here, we consider the following $(3+1)$-dimensional action of EYM-Massive
gravity with BI NED for the model
\begin{equation}
\mathcal{I}_{G}=-\frac{1}{16\pi }\int_{\mathcal{M}}d^{3+1}x\sqrt{-g}\left(
\mathcal{R}-2\Lambda +\mathcal{L}_{BI}(\mathcal{F}_{M})-\mathcal{F}%
_{YM}+m^{2}\sum_{i}c_{i}\mathcal{U}_{i}(g,f)\right) ,  \label{Action}
\end{equation}%
where $\mathcal{L}_{BI}(\mathcal{F}_{M})$ and $\mathcal{F}_{YM}=\mathbf{Tr}%
\left( F_{\mu \nu }^{\left( a\right) }F^{(a)\mu \nu }\right) $ are,
respectively, the Lagrangian of BI NED (\ref{BI}) and the YM invariant. In
addition, $m$ is related to the graviton mass while $f$ refers to an
auxiliary reference metric which its components depend on the metric under
consideration. Moreover, $c_{i}$'s are some free constants and $\mathcal{U}%
_{i}$'s are symmetric polynomials of the eigenvalues of $4\times 4$ matrix $%
\mathcal{K}_{\nu }^{\mu }=\sqrt{g^{\mu \sigma }f_{\sigma \nu }}$ which have
the following forms
\begin{eqnarray}
\mathcal{U}_{1} &=&\left[ \mathcal{K}\right] ,  \notag \\
\mathcal{U}_{2} &=&\left[ \mathcal{K}\right] ^{2}-\left[ \mathcal{K}^{2}%
\right] ,  \notag \\
\mathcal{U}_{3} &=&\left[ \mathcal{K}\right] ^{3}-3\left[ \mathcal{K}\right] %
\left[ \mathcal{K}^{2}\right] +2\left[ \mathcal{K}^{3}\right] ,  \notag \\
\mathcal{U}_{4} &=&\left[ \mathcal{K}\right] ^{4}-6\left[ \mathcal{K}^{2}%
\right] \left[ \mathcal{K}\right] ^{2}+8\left[ \mathcal{K}^{3}\right] \left[
\mathcal{K}\right] +3\left[ \mathcal{K}^{2}\right] ^{2}-6\left[ \mathcal{K}%
^{4}\right] ,  \notag \\
&&.  \notag \\
&&.  \notag \\
&&.  \notag
\end{eqnarray}%
where the rectangular bracket represents the trace of $\mathcal{K}_{\nu
}^{\mu }$. It is easy to obtain three tensorial field equations which come
from the variation of action (\ref{Action}) with respect to the metric
tensor $g_{\mu \nu }$, and the gauge potentials $A_{\mu }$ and $A_{\mu
}^{\left( a\right) }$ as
\begin{equation}
G_{\mu \nu }+\Lambda g_{\mu \nu }=T_{\mu \nu }^{M}+T_{\mu \nu
}^{YM}-m^{2}\chi _{\mu \nu },  \label{Field equation}
\end{equation}%
\begin{equation}
\partial _{\mu }\left[ \sqrt{-g}F^{\mu \nu }\partial _{\mathcal{F}}\mathcal{L%
}_{BI}(\mathcal{F})\right] =0,  \label{Maxwell equation}
\end{equation}%
\begin{equation}
\overset{\wedge }{D_{\mu }}F^{(a)\mu \nu }=0,  \label{YM equation}
\end{equation}%
where $\overset{\wedge }{D_{\mu }}$ is the covariant derivative of the gauge
field. The energy-momentum tensor of electromagnetic and YM fields, and
also, $\chi _{\mu \nu }$ can be written as
\begin{equation}
T_{\mu \nu }^{M}=\frac{1}{2}g_{\mu \nu }\mathcal{L}_{BI}(\mathcal{F}%
)-2F_{\mu \lambda }F_{\nu }^{\lambda }\partial _{\mathcal{F}}\mathcal{L}%
_{BI}(\mathcal{F}),  \label{Mmoment}
\end{equation}%
\begin{equation}
T_{\mu \nu }^{YM}=-\frac{1}{2}g_{\mu \nu }F_{\rho \sigma }^{(a)}F^{(a)\rho
\sigma }+2F_{\mu \lambda }^{(a)}F_{\nu }^{(a)\lambda },  \label{YMmoment}
\end{equation}

\begin{eqnarray}
\chi _{\mu \nu } &=&-\frac{c_{1}}{2}\left( \mathcal{U}_{1}g_{\mu \nu }-%
\mathcal{K}_{\mu \nu }\right) -\frac{c_{2}}{2}\left( \mathcal{U}_{2}g_{\mu
\nu }-2\mathcal{U}_{1}\mathcal{K}_{\mu \nu }+2\mathcal{K}_{\mu \nu
}^{2}\right) -\frac{c_{3}}{2}(\mathcal{U}_{3}g_{\mu \nu }-3\mathcal{U}_{2}%
\mathcal{K}_{\mu \nu }+  \notag \\
&&6\mathcal{U}_{1}\mathcal{K}_{\mu \nu }^{2}-6\mathcal{K}_{\mu \nu }^{3})-%
\frac{c_{4}}{2}(\mathcal{U}_{4}g_{\mu \nu }-4\mathcal{U}_{3}\mathcal{K}_{\mu
\nu }+12\mathcal{U}_{2}\mathcal{K}_{\mu \nu }^{2}-24\mathcal{U}_{1}\mathcal{K%
}_{\mu \nu }^{3}+24\mathcal{K}_{\mu \nu }^{4})+ ... .
\end{eqnarray}

In addition, the YM tensor $F_{\mu \nu }^{\left( a\right) }$ has the
following form
\begin{equation}
F_{\mu \nu }^{(a)}=2\nabla _{\lbrack \mu }A_{\nu
]}^{(a)}+f_{(b)(c)}^{(a)}A_{\mu }^{(b)}A_{\nu }^{(c)},  \label{YMfield}
\end{equation}%
in which $A_{\mu }^{(a)}$\ is the YM potential and the symbols $%
f_{(b)(c)}^{(a)}$'s denote the real structure constants of the $3$%
-parameters YM gauge group $SU(2)$ (note: the structure constants can be
calculated by using the commutation relation of the gauge group generators).

In order to obtain the spherically symmetric black hole solutions of
EYM-Massive theory coupled to BI NED, we restrict attention to the following
metric
\begin{equation}
g_{\mu \nu }=diag\left[-f(r),f^{-1}(r),r^{2},r^{2}\sin ^{2}\theta \right],
\label{metric}
\end{equation}
with the following reference metric ansatz \cite{Vegh}
\begin{equation}
f_{\mu \nu }=diag\left[0,0,c^{2},c^{2}\sin ^{2}\theta \right],  \label{f11}
\end{equation}
where $c$ is an arbitrary positive constant. Using the metric ansatz (\ref%
{f11}), $\mathcal{U}_{i}$'s reduce to the following explicit forms \cite%
{Vegh}
\begin{equation}
\mathcal{U}_{1}=2cr^{-1},\ \ \ \mathcal{U}_{2}=2c^{2}r^{-2},\ \ \ \mathcal{U}%
_{i}=0 \;\;for\;\; i \geq 3.
\end{equation}

Considering the field equations (\ref{Maxwell equation}) with the following
radial gauge potential ansatz
\begin{equation}
A_{\mu }=h\left( r\right) \delta _{\mu }^{t},  \label{potential ansatz}
\end{equation}%
one can obtain the following differential equation
\begin{equation}
\beta ^{2}rE^{\prime }(r)+2E(r)\left[ \beta ^{2}-E^{2}(r)\right] =0,
\label{hdiff}
\end{equation}%
where $E(r)=-h^{\prime }(r)$ and prime refers to $d/dr$. Solving Eq. (\ref%
{hdiff}), we obtain
\begin{equation}
E(r)=\frac{q}{r^{2}}\left( 1+\frac{q^{2}}{\beta ^{2}r^{4}}\right) ^{-1/2},
\label{E1(r)}
\end{equation}%
where $q$ is an integration constant which is related to the total electric
charge of the black hole. It is clear that in the limit $\beta \rightarrow
\infty $, Eq. (\ref{E1(r)}) tends to $q/r^{2}$, and therefore, the Maxwell
electric field will be recovered.

Hereafter and for the sake of simplicity, we use the position dependent
generators $\mathbf{t}_{(r)}$, $\mathbf{t}_{(\theta )}$, and $\mathbf{t}%
_{(\varphi )}$ of the gauge group instead of the standard generators $%
\mathbf{t}_{(1)}$, $\mathbf{t}_{(2)}$, and $\mathbf{t}_{(3)}$. The relation
between the basis of $SU(2)$ group and the standard basis are
\begin{equation}
\begin{array}{c}
\mathbf{t}_{(r)}=\sin \theta \cos \nu \varphi \mathbf{t}_{(1)}+\sin \theta
\sin \nu \varphi \mathbf{t}_{(2)}+\cos \theta \mathbf{t}_{(3)} \\
\mathbf{t}_{(\theta )}=\cos \theta \cos \nu \varphi \mathbf{t}_{(1)}+\cos
\theta \sin \nu \varphi \mathbf{t}_{(2)}-\sin \theta \mathbf{t}_{(3)} \\
\mathbf{t}_{(\varphi )}=-\sin \nu \varphi \mathbf{t}_{(1)}+\cos \nu \varphi
\mathbf{t}_{(2)}%
\end{array}
,
\end{equation}%
and it is straightforward to show that these generators satisfy the
following commutation relations
\begin{equation}
\left[ \mathbf{t}_{(r)},\mathbf{t}_{(\theta )}\right] =\mathbf{t}_{(\varphi
)},\ \ \ \left[ \mathbf{t}_{(\varphi )},\mathbf{t}_{(r)}\right] =\mathbf{t}%
_{(\theta )},\ \ \ \left[ \mathbf{t}_{(\theta )},\mathbf{t}_{(\varphi )}%
\right] =\mathbf{t}_{(r)}.
\end{equation}

In order to solve the YM field equations (\ref{YM equation}), just like the
electromagnetic case, it is required to choose a gauge potential ansatz.
Here, we are interested in the magnetic Wu-Yang ansatz of the gauge
potential with the following nonzero components \cite%
{BalakinLemos,BalakinZayatsPLB}
\begin{equation}
A_{\theta }^{(a)}=\delta _{(\varphi )}^{(a)},\ \ \ A_{\varphi }^{(a)}=-\nu
\sin \theta \delta _{(\theta )}^{(a)},  \label{YMgauge}
\end{equation}%
where the magnetic parameter $\nu $ is a non-vanishing integer. It is easy
to show that the chosen Wu-Yang gauge potential (\ref{YMgauge}) satisfies
the YM field equations (\ref{YM equation}). Using the YM tensor field (\ref%
{YMfield}) with Wu-Yang ansatz (\ref{YMgauge}), one can show that the only
non-vanishing component of the YM field is
\begin{equation}
F_{\theta \varphi }^{(r)}=\nu \sin \theta .  \label{Ftp}
\end{equation}

Considering the metric (\ref{metric})\ with the electromagnetic (\ref{E1(r)}%
) and YM fields (\ref{Ftp}), one can show that the only two different
components of the field equations (\ref{Field equation}) are
\begin{eqnarray}
tt-component: e_{tt} &=&rf^{\prime }(r)+f(r)-1+\left( \Lambda -2\beta
^{2}\right) r^{2}-m^{2}\left( cc_{1}r+c^{2}c_{2}\right) +\frac{\nu ^{2}}{%
r^{2}}+2\beta \sqrt{q^{2}+\beta ^{2}r^{4}}=0,  \label{e1} \\
\theta \theta -component: e_{\theta \theta} &=&\frac{r}{2}f^{\prime \prime
}(r)+f^{\prime }(r)+\left( \Lambda -2\beta ^{2}\right) r-\frac{m^{2}}{2}%
cc_{1}-\frac{\nu ^{2}}{r^{3}}+\frac{2\beta ^{3}r^{3}}{\sqrt{q^{2}+\beta
^{2}r^{4}}}=0,  \label{e2}
\end{eqnarray}

Since there is one common unknown function in both $e_{tt}$ and $e_{\theta
\theta }$ equations, it is expected to find that the mentioned field
equations are not independent. After some manipulations, one can obtain the
second order field equation by a suitable combination of first order one as
\begin{equation}
e_{\theta \theta }=e_{tt}^{\prime }+\frac{1}{r}e_{tt}
\end{equation}%
and therefore, the solutions of $e_{tt}$ with an integration constant
satisfy $e_{\theta \theta }$ equation, directly. Solving Eq. (\ref{e1}), we
can obtain the following metric function
\begin{equation}
f(r)=1-\frac{m_{0}}{r}-\frac{\Lambda r^{2}}{3}+\frac{\nu ^{2}}{r^{2}}+\frac{%
m^{2}}{2r}\left( cc_{1}r^{2}+2c^{2}c_{2}r\right) +\frac{2\beta ^{2}r^{2}}{3}%
\left( 1-\mathcal{H}_{1}\right) ,  \label{metric function}
\end{equation}%
where $\mathcal{H}_{1}={}_{2}F_{1}\left( -\frac{1}{2},-\frac{3}{4};\frac{1}{4%
};-\frac{q^{2}}{\beta ^{2}r^{4}}\right) $ is a hypergeometric function and $%
m_{0}$ is the only integration constant which is related to the total mass
of black hole. Considering the obtained $f(r)$, one finds that the fourth
term is related to the magnetic charge (hair), the fifth term is related to
the massive gravitons, and finally, the last term comes from the
nonlinearity of electric charge. Now, it is worthwhile to investigate the
asymptotic behavior of the nonlinearity parameter $\beta $ on the solutions.
Expanding the metric function (\ref{metric function}) for large $\beta $
\begin{equation}
f(r)=1-\frac{m_{0}}{r}-\frac{\Lambda r^{2}}{3}+\frac{\nu ^{2}}{r^{2}}+\frac{%
m^{2}}{2r}\left( cc_{1}r^{2}+2c^{2}c_{2}r\right) +\frac{q^{2}}{r^{2}}-\frac{%
q^{4}}{20\beta ^{2}r^{6}}+\mathcal{O}\left( \frac{1}{\beta ^{4}}\right) ,
\label{largeB}
\end{equation}%
one can recover the Maxwellian limit of the solutions. Therefore, for the
massless graviton, $m=0$, and linear electrodynamics, $\beta \rightarrow
\infty $, the metric function (\ref{largeB}) reduces to the EYM solution
with Maxwell field, as we expected. On the other hand, for small values of
the nonlinearity parameter (highly nonlinear solutions), we have
\begin{equation}
f(r)=1-\frac{m_{0}}{r}-\frac{\Lambda r^{2}}{3}+\frac{\nu ^{2}}{r^{2}}+\frac{%
m^{2}}{2r}\left( cc_{1}r^{2}+2c^{2}c_{2}r\right) +\frac{\Gamma ^{2}\left(
1/4\right) }{3}\sqrt{\frac{\beta }{\pi }}\frac{q^{3/2}}{r}+\mathcal{O}\left(
\beta \right) ,  \label{smallB}
\end{equation}%
which shows that the black hole is neutral at the highly nonlinear regime ($%
\beta \rightarrow 0$).

Considering Eq. (\ref{metric function}), it is clear that the asymptotical
behavior of the solutions is adS (or dS) provided $\Lambda <0$ (or $\Lambda
>0$). In order to find the singularity of the solutions, one can obtain the
Kretschmann scalar as
\begin{equation}
R_{\mu \nu \lambda \kappa }R^{\mu \nu \lambda \kappa }=\frac{4}{r^{4}}\left[
1+f^{2}(r)-2f(r)+\left[ rf^{\prime }(r)\right] ^{2}+\left( \frac{%
r^{2}f^{\prime \prime }(r)}{2}\right) ^{2}\right] ,  \label{kr}
\end{equation}%
which by inserting (\ref{metric function}), it is straightforward to show
that the Kretschmann scalar has the following behavior
\begin{equation}
\lim_{r\rightarrow 0}\left( R_{\mu \nu \lambda \kappa }R^{\mu \nu \lambda
\kappa }\right) =\infty ,\ \ \ \ \lim_{r\rightarrow \infty }\left( R_{\mu
\nu \lambda \kappa }R^{\mu \nu \lambda \kappa }\right) =\frac{8\Lambda ^{2}}{%
3}.  \label{RRbehavior}
\end{equation}

\begin{figure}[tbp]
$%
\begin{array}{c}
\epsfxsize=9cm \epsffile{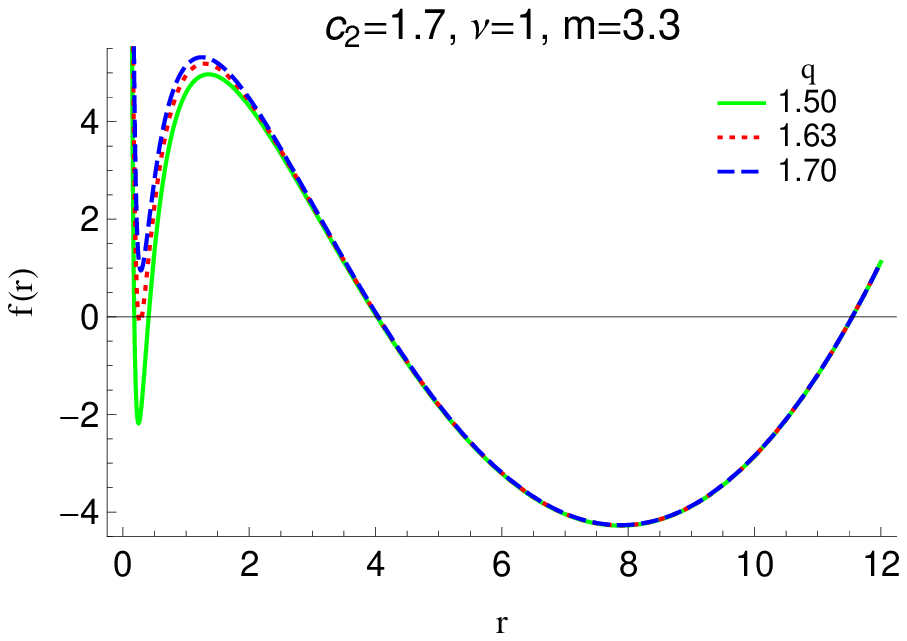} \\
\epsfxsize=9cm \epsffile{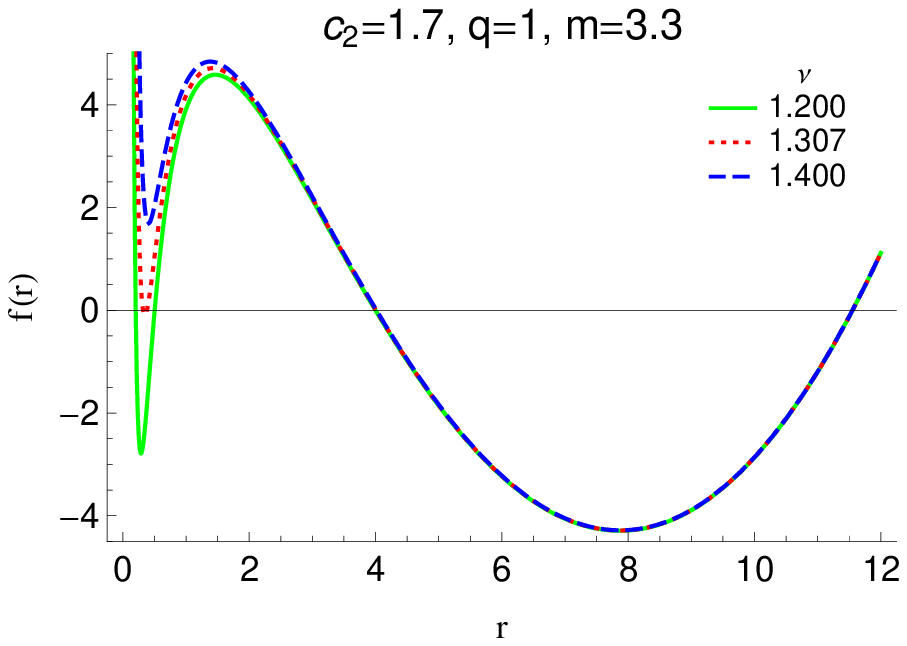} \\
\epsfxsize=9cm \epsffile{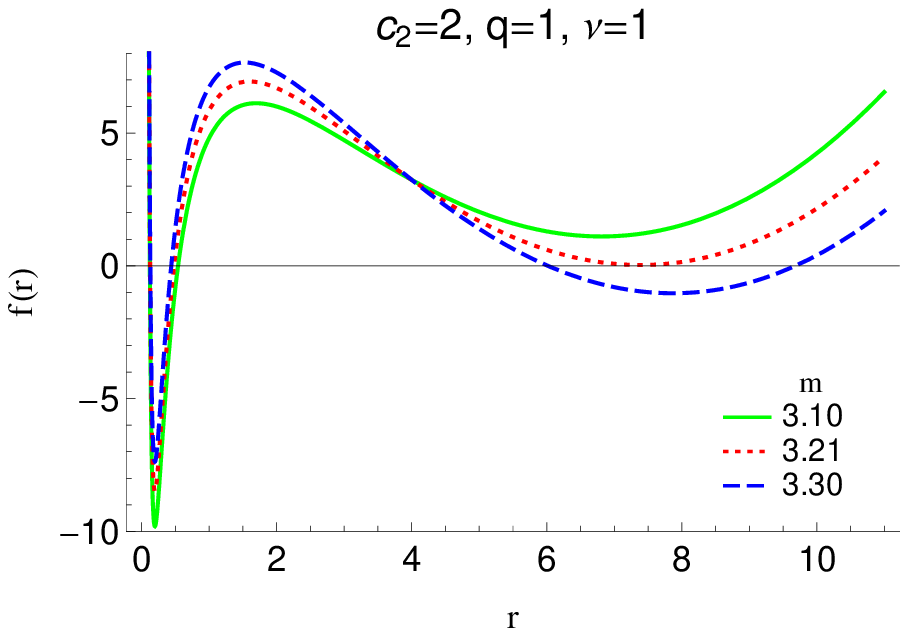}%
\end{array}
$%
\caption{ $f(r)$ versus $r$ for $\Lambda =-1$, $m_{0}=12.89$, $\protect\beta %
=1$, and $c=-c_{1}=1$.}
\label{metricfig}
\end{figure}

Equation (\ref{RRbehavior}) shows that there is an essential singularity
located at the origin, $r=0$. Moreover, the asymptotical behavior of the
Kretschmann scalar for the large enough $r$ confirms that the solutions are
asymptotically (a)dS. Moreover, this singularity can be covered with an
event horizon (for $\Lambda<0$), and therefore, one can interpret the
singularity as a black hole (Fig. \ref{metricfig}). As a final point of this
section, we should note that the metric function can possess more than two
real positive roots which this behavior is due to giving mass to the
gravitons (see \cite{HendiMassive,HendiMassiveBTZ} for more details).

\section{Thermodynamics \label{law}}

\subsection{Conserved and thermodynamic quantities \label{law}}

Here, we first obtain the conserved and thermodynamic quantities of the
black hole solutions, and then examine the validity of the first law of
thermodynamics.

The Hawking temperature of the black hole on the event (outermost) horizon, $%
r_{+}$, can be obtained by using the definition of surface gravity, $\kappa$%
,
\begin{equation}
T=\frac{\kappa }{2\pi }=\frac{1}{2\pi }\sqrt{-\frac{1}{2}\left( \nabla _{\mu
}\chi _{\nu }\right) \left( \nabla ^{\mu }\chi ^{\nu }\right) },  \label{SG}
\end{equation}%
where $\chi =\partial_{t}$ is the null Killing vector of the horizon. Thus,
the temperature is obtained as
\begin{equation}
T=\left. \frac{f^{\prime }(r)}{4\pi }\right\vert _{r=r_{+}}=\frac{1}{4\pi
r_{+}}\left[ 1-\Lambda r_{+}^{2}-\frac{\nu ^{2}}{r_{+}^{2}}+m^{2}\left(
cc_{1}r_{+}+c^{2}c_{2}\right) +2\beta ^{2}r_{+}^{2}\left( 1-\sqrt{1+\frac{%
q^{2}}{\beta ^{2}r_{+}^{4}}}\right) \right] .  \label{T+}
\end{equation}

It is worthwhile to mention that fourth term of RHS of Eq. (\ref{T+}) does
not depend on the horizon radius, and therefore, one can regard it as a
constant background temperature, $T_{0}=\frac{m^{2}cc_{1}}{4\pi}$. As a
result, we can investigate the solutions by using an effective temperature, $%
\hat{T}=T-T_{0}$.

The electric potential $\Phi $, measured at infinity with respect to the
horizon $r_{+}$, is obtained by
\begin{equation}
\Phi _{E}=\left. A_{\mu }\chi ^{\mu }\right\vert _{r\rightarrow \infty
}-\left. A_{\mu }\chi ^{\mu }\right\vert _{r=r_{+}}=\frac{q}{r_{+}}%
\;{}_{2}F_{1}\left( \frac{1}{2},\frac{1}{4};\frac{5}{4};-\frac{q^{2}}{\beta
^{2}r_{+}^{4}}\right).  \label{potential}
\end{equation}

Since we are working in the context of Einstein gravity, the entropy of the
black holes still obeys the so-called area law. Therefore, the entropy of
black holes is equal to one-quarter of the horizon area with the following
explicit form
\begin{equation}
S=\pi r_{+}^{2}.  \label{entropy}
\end{equation}

In order to obtain the electric charge of the black hole, we use the flux of
the electric field at infinity, yielding
\begin{equation}
Q_{E}=q.  \label{Mcharge}
\end{equation}

It was shown that by using the Hamiltonian approach, one can obtain the
total mass $M$ in the context of massive gravity as \cite{CaiMassive}
\begin{equation}
M=\frac{m_{0}}{2},  \label{mass}
\end{equation}
where $m_{0}$ comes from the fact that $f(r=r_{+})=0$.


In the limit that the parameter $\beta $ is small, one may think that the
last term in Eq. (\ref{smallB}) should contribute to the total mass along
with $m_{0}$ (since they have the same radial dependence). But it is not
correct since in order to calculate the total energy (mass) of spacetime, we
have to regard the related action for large values of $r$. It is notable
that the asymptotic behavior of the metric function for both limits $%
r\longrightarrow \infty $ and $\beta \longrightarrow \infty $ is the same.
So, we should consider Eq. (\ref{largeB}) to find the mass term, as a
coefficient of $r^{-1}$ in four dimensions. However, the functional form of
the mass term should be proportional to $r^{-1}$ for arbitrary values of $r$.

We should also note that the total mass is related to the geometrical mass,
which is an integration constant of the gravitational field equation. If we
regard the last term of Eq. (\ref{smallB}) as a (piece of) mass-term, two
problems are appeared; the first one is related to the higher-order series
expansion of Eq. (\ref{smallB}), in which they will be related to higher
orders of mass with the same dimensional analysis, but they are not
proportional to $r^{-1}$ at all. The second one is related to the Smarr
relation. It is straightforward to check that the Smarr relation is valid
only for the mass related to the geometrical mass ($m_{0}$). Regarding the
mentioned additional term, the Smarr relation is violated.

In addition, since the considered gravitational configuration has a
time-like Killing vector, it is straightforward to calculate the energy
(mass) of the system as the corresponding conserved quantity. Using the
Arnowitt-Deser-Misner (ADM) method \cite{A1,A2}, one finds the mentioned
conserved quantity for general solutions, $f(r)$, is related to the
geometrical mass, $m_{0}$.


Now, we are in a position to check the validity of the first law of
thermodynamics. To do so, we use the entropy (\ref{entropy}), the electric
charge (\ref{Mcharge}), and the mass (\ref{mass}) to obtain mass as a
function of entropy and electric charge
\begin{equation}
M\left( S,Q_{E}\right) =\frac{1}{2}\left( \frac{S}{\pi }\right) ^{3/2}\left[
\frac{\pi }{S}-\frac{\Lambda }{3}+\left( \frac{\pi \nu }{S}\right) ^{2}+%
\frac{2\beta ^{2}}{3}\left( 1-\mathcal{H}_{3}\right) \right] +\frac{m^{2}}{%
4\pi }\left( cc_{1}S+2c^{2}c_{2}\sqrt{\frac{S}{\pi }}\right) ,  \label{smarr}
\end{equation}
where $\mathcal{H}_{3}={}_{2}F_{1}\left( -\frac{1}{2},-\frac{3}{4};\frac{1}{4%
};-\left( \frac{\pi Q_{E}}{\beta S}\right) ^{2}\right) $. We consider the
entropy ($S$) and electric charge ($Q_{E}$)\ as a complete set of extensive
parameters, and define the temperature ($T$)\ and electric potential ($\Phi
_{E}$) as the intensive parameters conjugate to them
\begin{equation}
T=\left( \frac{\partial M}{\partial S}\right) _{Q_{E}}=\frac{1}{4\pi }\sqrt{%
\frac{\pi }{S}}\left[ 1-\Lambda S-\frac{\pi \nu ^{2}}{S}+\frac{2\beta ^{2}S}{%
\pi }\left( 1-\mathcal{H}_{3}\right) -\frac{4\beta ^{2}S^{2}}{3\pi } \left(%
\frac{d\mathcal{H}_{3}}{dS}\right) _{Q_{E}} +m^{2}\left( cc_{1}\sqrt{\frac{S%
}{\pi }}+c^{2}c_{2}\right) \right] ,  \label{1}
\end{equation}
\begin{equation}
\Phi _{E}=\left( \frac{\partial M}{\partial Q_{E}}\right) _{S}=\sqrt{\frac{%
\pi }{S}}Q_{E}\;{}_2F_{1}\left( \frac{1}{2},\frac{1}{4};\frac{5}{4};-\left(
\frac{\pi Q_{E} }{\beta S}\right) ^{2}\right) .  \label{2}
\end{equation}

Using Eqs. (\ref{entropy}) and (\ref{Mcharge}), one can easily show that the
temperature (\ref{1}) and electric potential (\ref{2}) are, respectively,
equal to Eqs. (\ref{T+}) and (\ref{potential}). Thus, these quantities
satisfy the first law of thermodynamics
\begin{equation}
dM=TdS+\Phi _{E}dQ_{E}.  \label{firstlaw}
\end{equation}

On the other hand, the obtained black holes enjoy a global YM charge as
well. In order to find this magnetic charge, we use the following definition
\begin{equation}
Q_{YM}=\frac{1}{4\pi }\int \sqrt{F_{\theta \varphi }^{(a)}F_{\theta \varphi
}^{(a)}}d\theta d\varphi =\nu .  \label{YMcharge}
\end{equation}

In order to complete the first law of thermodynamics in differential form (%
\ref{firstlaw}), one can consider the YM charge as an extensive
thermodynamic variable and introduce an effective YM potential conjugate to
it as an intensive variable
\begin{equation}
\Phi _{YM}=\left( \frac{\partial M}{\partial Q_{YM}}\right)
_{S,Q_{E}}=\left( \frac{\partial M}{\partial \nu }\right) _{S,Q_{E}}\left/
\left( \frac{\partial Q_{YM}}{\partial \nu }\right) _{S,Q_{E}}\right. =\nu
\sqrt{\frac{\pi }{S}}=\frac{\nu }{r_{+}},
\end{equation}%
which satisfies the first law of thermodynamics in a more complete way
\begin{equation}
dM=TdS+\Phi _{E}dQ_{E}+\Phi _{YM}dQ_{YM}.
\end{equation}

Regarding the differential form of the first law, it is worth mentioning
that this equation may be completed by other additional terms, such as $VdP$
in the extended phase space. In order to check the validity of the existence
of such terms, one should check the first law in a non-differential form,
the so-called Smarr relation. After some manipulations, one can find that
\begin{equation}
M=2TS+\Phi _{E}Q_{E}+\Phi _{YM}Q_{YM}-2VP-\mathcal{B}\beta -\mathcal{C}c_{1},
\label{Smarr}
\end{equation}%
where%
\begin{equation}
P=-\frac{\Lambda }{8\pi },\ \ V=\left( \frac{\partial M}{\partial P}\right)
_{S,Q_{E},Q_{YM},\beta ,c_{1}},\ \ \mathcal{B}=\left( \frac{\partial M}{%
\partial \beta }\right) _{S,Q_{E},Q_{YM},P,c_{1}},\ \ \mathcal{C}=\left(
\frac{\partial M}{\partial c_{1}}\right) _{S,Q_{E},Q_{YM},P,\beta },
\end{equation}%
which confirm that the existence of additional terms and leads to a more
complete form of the first law of thermodynamics%
\begin{equation}
dM=TdS+\Phi _{E}dQ_{E}+\Phi _{YM}dQ_{YM}+VdP+\mathcal{B}d\beta +\mathcal{C}%
dc_{1}.  \label{FL}
\end{equation}

It is worthwhile to mention that although it is possible to add $C_{2}dc_{2}$%
\ to the first law of thermodynamics (\ref{FL}) mathematically, we are not
allowed due to the fact that all intensive and extensive thermodynamic
parameters should appear in the Smarr formula (\ref{Smarr}). Therefore, we
considered $c_{2}$\ as a constant (not a thermodynamic variable) since it
did not appear in the Smarr formula.

\subsection{Thermal stability \label{thermal}}

In this section, we use the heat capacity for investigating the thermal
stability of the obtained black hole solutions. In this regard, one should
consider the sign of heat capacity (its positivity and negativity) to study
the stability conditions. The root of heat capacity (or temperature)
represents a bound point. This point is a kind of border which is located
between physical black holes related to the positive temperature and
non-physical ones with a negative temperature. On the other hand, in our
case, both divergence points of the heat capacity indicate one thermal phase
transition point where black holes jump from one divergency to the other
one. Besides, the heat capacity changes sign at such divergence points. So,
one can conclude that the divergence point is a kind of bound-like point
which is located between unstable black holes with negative heat capacity
and stable (or metastable) ones. Therefore, it is logical to say that the
physical stable black holes are located everywhere that both the heat
capacity and temperature are positive, simultaneously.

Here, we study the thermal stability of the asymptotically adS solutions
with $\Lambda <0$. The heat capacity at constant electric and YM charges is
given by
\begin{equation}
C_{Q_{E},Q_{YM}}=\frac{T}{\left( \frac{\partial ^{2}M}{\partial S^{2}}%
\right) _{Q_{E},Q_{YM}}},  \label{heat}
\end{equation}%
where $T$ has been obtained in Eq. (\ref{T+}). Considering (\ref{entropy}), (%
\ref{Mcharge}) and (\ref{smarr}), one can easily show that the denominator
of heat capacity is
\begin{equation}
\left( \frac{\partial ^{2}M}{\partial S^{2}}\right) _{Q_{E},Q_{YM}}=\frac{1}{%
8\pi ^{2}r_{+}^{3}}\left[ \left( \beta ^{2}-\Lambda \right)
r_{+}^{2}-1-m^{2}c^{2}c_{2}+\frac{3\nu ^{2}}{r_{+}^{2}}+2\beta q\left( 1+%
\frac{\beta ^{2}r_{+}^{4}}{q^{2}}\right) ^{-1/2}\left( 1-\frac{\beta
^{2}r_{+}^{4}}{q^{2}}\right) \right] .  \label{denom}
\end{equation}

We recall that thermal stability criteria are based on the sign of heat
capacity and it may change at root and divergence points. Therefore, it is
necessary to look for the root and divergence points of the heat capacity at
the first step. But unfortunately, because of the complexity of Eq. (\ref%
{heat}), it is not possible to obtain the root and divergencies of the heat
capacity, analytically. So, we adopt the numerical analysis to obtain both
bound and thermal phase transition points.

Before applying the numerical calculations, we are interested to clarify the
general behavior of the heat capacity and temperature for the small and
large black holes. For the fixed values of different parameters, there could
exist two special $r_{+}$'s, say $r_{+\min }$ and $r_{+\max }$ (see Fig. \ref%
{CQadS}). The small black holes and large black holes are located before $%
r_{+\min }$ and after $r_{+\max }$, respectively. The region of $r_{+\min
}<r_{+}<r_{+\max }$ belongs to the intermediate black holes. Using the
series expanding of (\ref{heat}), one obtains
\begin{equation}
\left\{
\begin{array}{c}
C_{Q_{E},Q_{YM}}=-\frac{2\pi }{3}r_{+}^{2}+\mathcal{O}\left( r_{+}^{4}\right)
\\
T=-\frac{\nu ^{2}}{4\pi r_{+}^{3}}+\mathcal{O}\left( \frac{1}{r_{+}}\right)%
\end{array}%
\right. ,\ \ \ \ \ \text{for\ small\ }r_{+},  \label{H1}
\end{equation}%
\begin{equation}
\left\{
\begin{array}{c}
C_{Q_{E},Q_{YM}}=Const.+2\pi r_{+}^{2}+\mathcal{O}\left( r_{+}\right) \\
T=Const.-\frac{\Lambda r_{+}}{4\pi }+\mathcal{O}\left( \frac{1}{r_{+}}\right)%
\end{array}%
\right. ,\ \ \ \ \ \text{for\ large}\ r_{+}.  \label{H2}
\end{equation}

\begin{figure}[tbp]
$%
\begin{array}{ccc}
\epsfxsize=7.5cm \epsffile{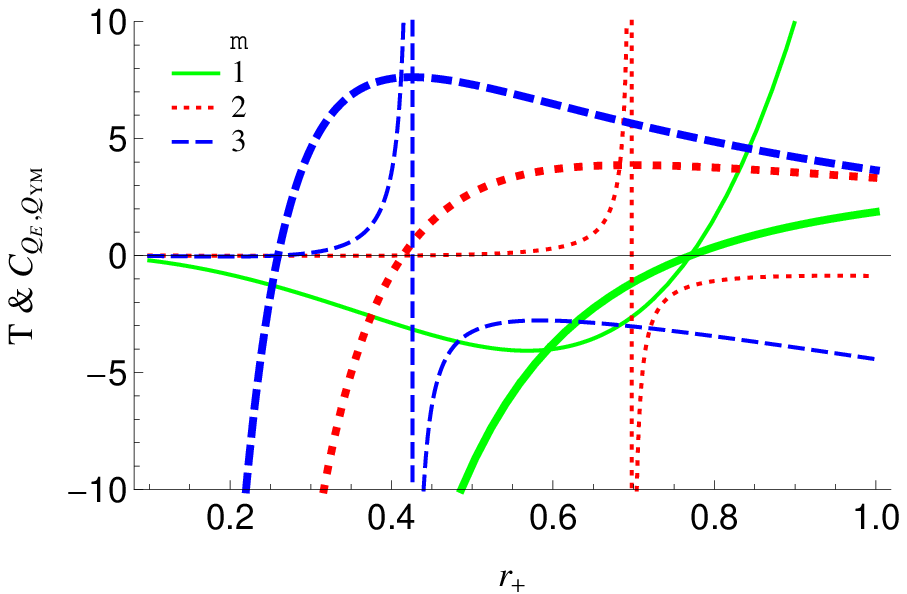} & \epsfxsize=7.5cm \epsffile{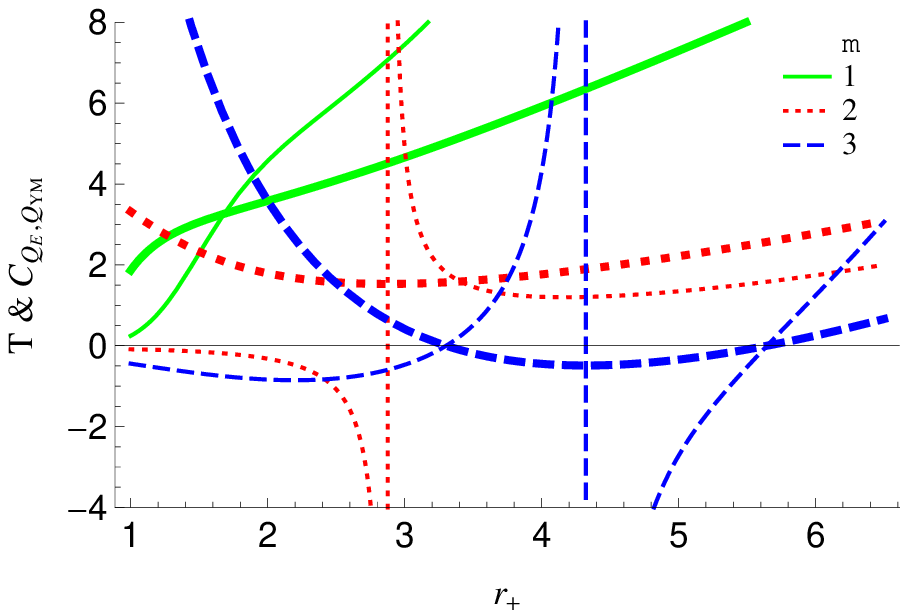} &
\end{array}
$%
\caption{ $C_{Q_{E},Q_{YM}}$ (thin lines) and $T$ (bold lines) versus $r_{+}$
for $\Lambda =c_{1}=-1$, $q=\protect\beta =\protect\nu =c=1$, and $c_{2}=2 $%
. }
\label{CQadS}
\end{figure}

Considering Eq. (\ref{H1}), it is clear that for sufficiently small $r_{+}$,
the heat capacity and temperature are negative, and therefore, we have an
unstable and non-physical black hole. Whereas from Eq. (\ref{H2}), we find
that for large $r_{+}$, both heat capacity and temperature are positive and
there exists stable and physical black hole. In other words, Eqs. (\ref{H1})
and (\ref{H2}) confirm that the small black holes ($r_{+}<r_{+\min }$) are
unstable and non-physical, whereas the large black holes ($r_{+}>r_{+\max }$%
) are physical and enjoy thermal stability. It is notable that in a special
case there is just one specific horizon radius, $r_{+s}$. In this case, we
have unstable black holes for $r_{+}<r_{+s}$ and stable ones for $%
r_{+}>r_{+s}$. However, it is not possible to identify this last property
analytically, but we show it in Fig. \ref{CQadS} (see continues line).

Now, we back to the numerical analysis of the heat capacity. Although we
studied the general behavior of the heat capacity for the small and large
black holes, the numerical calculations help us to classified the
intermediate black holes ($r_{+\min }<r_{+}<r_{+\max }$). However, we are
not going to study all possible behaviors of the heat capacity (because they
contain different cases due to lots of free parameters) and just take some
interesting ones.

Figure \ref{CQadS} shows some different possibilities for the heat capacity.
Clearly, this figure confirms that the small black holes are unstable (Eq. (%
\ref{H1})) and large black holes are stable (Eq. (\ref{H2})). According to
the numerical analysis, we find that the heat capacity contains ($i$) only
one bound point, ($ii$) one bound point and two divergencies, and ($iii$)
three bound points and two divergencies. In the first case, we have unstable
and non-physical black holes before the bound point ($r_{+s}$), but after
this point, stable and physical black holes are presented. It is worthwhile
to recall that from Eqs. (\ref{H1}) and (\ref{H2}), we expected such
behavior. In the second case, we have stable and physical solutions between
the bound point and smaller divergency. There are physical and unstable
black holes between two divergencies. It is notable to mention that the
large black holes are stable and physical as well. As for the last case,
there are stable and unstable solutions respectively before and after the
larger divergency.

\begin{center}
\begin{tabular}{ccccccccc}
\hline\hline
$m$ &  & $\beta $ &  & $q$ &  & $\nu $ &  & $r_{+crit}$ \\ \hline\hline
$1.0$ &  & $1.0$ &  & $1.0$ &  & $1.0$ &  & $0.7705$ \\ \hline
$1.1$ &  & $1.0$ &  & $1.0$ &  & $1.0$ &  & $0.7311$ \\ \hline
$1.2$ &  & $1.0$ &  & $1.0$ &  & $1.0$ &  & $0.6901$ \\ \hline
$1.0$ &  & $2.0$ &  & $1.0$ &  & $1.0$ &  & $0.8141$ \\ \hline
$1.0$ &  & $3.0$ &  & $1.0$ &  & $1.0$ &  & $0.8255$ \\ \hline
$1.0$ &  & $1.0$ &  & $2.0$ &  & $1.0$ &  & $1.0941$ \\ \hline
$1.0$ &  & $1.0$ &  & $3.0$ &  & $1.0$ &  & $1.4391$ \\ \hline
$1.0$ &  & $1.0$ &  & $1.0$ &  & $2.0$ &  & $1.2243$ \\ \hline
$1.0$ &  & $1.0$ &  & $1.0$ &  & $3.0$ &  & $1.5904$ \\ \hline
\end{tabular}

Table $I$: case ($i$): The root of heat capacity for $\Lambda =-1$, $c=1$, $%
c_{1}=-1$, and $c_{2}=2$.

\begin{tabular}{ccccccccccc}
\hline\hline
$m$ &  & $\beta $ &  & $q$ &  & $\nu $ &  & $r_{+\min }$ & $%
\begin{array}{c}
\text{smaller} \\
\text{divergency}%
\end{array}%
$ & $r_{+\max }$ \\ \hline\hline
$2.0$ &  & $1.0$ &  & $1.0$ &  & $1.0$ &  & $0.4142$ & $0.6974$ & $2.8772$
\\ \hline
$2.1$ &  & $1.0$ &  & $1.0$ &  & $1.0$ &  & $0.3913$ & $0.6530$ & $3.0278$
\\ \hline
$2.2$ &  & $1.0$ &  & $1.0$ &  & $1.0$ &  & $0.3706$ & $0.6146$ & $3.1760$
\\ \hline
$2.0$ &  & $2.0$ &  & $1.0$ &  & $1.0$ &  & $0.4602$ & $0.7918$ & $2.8768$
\\ \hline
$2.0$ &  & $5.0$ &  & $1.0$ &  & $1.0$ &  & $0.5111$ & $0.8412$ & $2.8766$
\\ \hline
$2.0$ &  & $1.0$ &  & $1.5$ &  & $1.0$ &  & $0.4580$ & $0.7872$ & $2.7861$
\\ \hline
$2.0$ &  & $1.0$ &  & $2.0$ &  & $1.0$ &  & $0.5203$ & $0.9384$ & $2.6284$
\\ \hline
$2.0$ &  & $1.0$ &  & $1.0$ &  & $1.5$ &  & $0.6391$ & $1.0627$ & $2.7830$
\\ \hline
$2.0$ &  & $1.0$ &  & $1.0$ &  & $2.0$ &  & $0.8702$ & $1.4623$ & $2.6076$
\\ \hline
\end{tabular}

Table $II$: case ($ii$): The root and divergencies of the heat capacity for $%
\Lambda =-1$, $c=1$, $c_{1}=-1$, and $c_{2}=2$.

\begin{tabular}{ccccccccccccc}
\hline\hline
$m$ &  & $\beta $ &  & $q$ &  & $\nu $ &  & $r_{+\min }$ & $%
\begin{array}{c}
\text{smaller} \\
\text{divergency}%
\end{array}%
$ & middle root & $%
\begin{array}{c}
\text{larger} \\
\text{divergency}%
\end{array}%
$ & $r_{+\max }$ \\ \hline\hline
$3.0$ &  & $1.0$ &  & $1.0$ &  & $1.0$ &  & $0.259388$ & $0.425718$ & $%
3.302860$ & $4.321927$ & $5.645748$ \\ \hline
$3.1$ &  & $1.0$ &  & $1.0$ &  & $1.0$ &  & $0.249991$ & $0.410556$ & $%
3.049203$ & $4.463070$ & $6.512390$ \\ \hline
$3.2$ &  & $1.0$ &  & $1.0$ &  & $1.0$ &  & $0.241256$ & $0.396511$ & $%
2.889876$ & $4.604034$ & $7.304464$ \\ \hline
$3.0$ &  & $2.0$ &  & $1.0$ &  & $1.0$ &  & $0.275692$ & $0.461745$ & $%
3.302797$ & $4.321903$ & $5.645750$ \\ \hline
$3.0$ &  & $5.0$ &  & $1.0$ &  & $1.0$ &  & $0.315215$ & $0.536320$ & $%
3.302779$ & $4.321897$ & $5.645751$ \\ \hline
$3.0$ &  & $1.0$ &  & $2.0$ &  & $1.0$ &  & $0.280515$ & $0.455785$ & $%
3.180750$ & $4.263744$ & $5.685150$ \\ \hline
$3.0$ &  & $1.0$ &  & $3.0$ &  & $1.0$ &  & $0.308521$ & $0.493284$ & $%
2.963691$ & $4.158349$ & $5.745924$ \\ \hline
$3.0$ &  & $1.0$ &  & $1.0$ &  & $2.0$ &  & $0.558740$ & $0.864659$ & $%
3.179227$ & $4.263212$ & $5.685196$ \\ \hline
$3.0$ &  & $1.0$ &  & $1.0$ &  & $3.0$ &  & $0.919670$ & $1.306670$ & $%
2.951630$ & $4.154812$ & $5.746137$ \\ \hline
\end{tabular}%
\\[0pt]

Table $III$: case ($iii$): The root and divergence points of the heat
capacity for $\Lambda =-1$, $c=1$, $c_{1}=-1$, and $c_{2}=2$.
\end{center}


In addition, we investigate the effects of different parameters on the bound
points and divergencies of the heat capacity in tables $I-III$. From the
table $I$, we find that the specific horizon radius, $r_{+s}$, increases as
the electric (magnetic) charge of black hole increases too. This could
happen when the black hole absorbs electric (magnetic) charge. As a result,
the region of unstable black holes increases. When the nonlinearity
parameter increases and the nonlinear theory tends to the Maxwell case (\ref%
{SerBI}), the critical horizon radius increases. On the contrary, $r_{+s}$
is a decreasing function of the graviton mass ($m$). So, by increasing $m$,
the region of unstable black holes decreases. Considering table $II$, it is
clear that the smaller root ($r_{+\min }$) and smaller divergency are
decreasing functions of $m$, but the larger divergency ($r_{+\max }$)
increases as the massive parameter increases. In addition, we have found the
same effects for $\beta $, $q$, and $\nu $, but opposite behavior is seen
for $m$. Table $III$ shows that the smaller root ($r_{+\min }$), the smaller
divergency, and middle root decrease as the massive parameter increases,
whereas the larger divergency and the larger root ($r_{+\max }$) are
increasing functions of $m$. Like case ($ii$), one can see the same behavior
for $\beta $, $q$, and $\nu $. The smaller divergency, $r_{+\min }$, and $%
r_{+\max }$ are increasing functions of these parameters ($m$, $\beta $, $q$%
, and $\nu $), but the middle and larger divergency are decreasing functions
of them. Based on these three tables, we conclude that the qualitative
effects of $\beta $, $q$, and $\nu $ on the heat capacity are quite the same.

\subsection{$P-V$ criticality in the extended phase space \label{PV}}

It is well known that the most black holes can undergo a van der Waals like
phase transition when one considers the cosmological constant as a
thermodynamic pressure. In this section, we employ this analogy between the
cosmological constant and pressure in the canonical ensemble (fixed $Q_{E}$,
$Q_{YM}$, $\beta $, and $c_{1}$) to investigate the $P-V$ criticality and
study phase transition of obtained black holes in extended phase space.
Using the temperature given in Eq. (\ref{T+}) and the relation of $%
P=-\Lambda /8\pi $, it is straightforward to show that the equation of state
is given by
\begin{equation}
P(r_{+},\hat{T})=\frac{\hat{T}}{2r_{+}}-\frac{1}{8\pi r_{+}^{2}}\left[ 1-%
\frac{\nu ^{2}}{r_{+}^{2}}+m^{2}c^{2}c_{2}+2\beta ^{2}r_{+}^{2}\left( 1-%
\sqrt{1+\frac{q^{2}}{\beta ^{2}r_{+}^{4}}}\right) \right] ,  \label{PP1}
\end{equation}%
where $\hat{T}=T-\frac{m^{2}cc_{1}}{4\pi }$ and we made this choice in order
to have a unique critical temperature (see appendix for more details). The
thermodynamic volume is an extensive parameter which is conjugated to the
pressure and has the following form
\begin{equation}
V=\left( \frac{\partial H}{\partial P}\right) _{S},  \label{V}
\end{equation}%
where $H$ is the enthalpy of the system. In this perspective, the total mass
of black hole plays the role of enthalpy instead of internal energy due to
the fact that the cosmological constant is not a fixed parameter anymore and
it is actually a thermodynamic variable. Therefore, the thermodynamic volume
is calculated as
\begin{equation}
V=\frac{4}{3}\pi r_{+}^{3}.  \label{VV}
\end{equation}

Hereafter, we use $r_{+}$ instead of $V$ as a thermodynamic variable since
it is proportional to the specific volume \cite%
{Hendi-Vahidinia,Kubiznak-MannJHEP}. In order to study the phase transition
of the black holes, we need to obtain the Gibbs free energy. In this
extended phase space, one can determine the Gibbs free energy by using the
following definition
\begin{equation}
G=H-TS=-\frac{2\pi r_{+}^{3}}{3}P+\frac{3\nu ^{2}}{4r_{+}}+\frac{r_{+}}{4}%
\left( 1+m^{2}c^{2}c_{2}\right) -\frac{\beta ^{2}r_{+}^{3}}{6}\left( 1+2%
\mathcal{H}_{1+}-3\sqrt{1+\frac{q^{2}}{\beta ^{2}r_{+}^{4}}}\right) ,
\label{G}
\end{equation}%
where $\mathcal{H}_{1+}=\mathcal{H}_{1}(r=r_{+})$. In addition, using the
properties of inflection point
\begin{equation}
\left( \frac{\partial P(r_{+},\hat{T})}{\partial r_{+}}\right) _{\hat{T}=%
\hat{T}_{c},r_{+}=r_{+c}}=\left( \frac{\partial ^{2}P(r_{+},\hat{T})}{%
\partial r_{+}^{2}}\right) _{\hat{T}=\hat{T}_{c},r_{+}=r_{+c}}=0,
\label{inflection point}
\end{equation}%
and after some manipulations, we obtain the following equation
\begin{equation}
\left( 1+m^{2}c^{2}c_{2}\right) r_{+c}^{2}-6\nu ^{2}-2q^{2}\left( 3+\frac{%
q^{2}}{\beta ^{2}r_{+c}^{4}}\right) \left( 1+\frac{q^{2}}{\beta
^{2}r_{+c}^{4}}\right) ^{-3/2}=0.  \label{crit}
\end{equation}

Considering this equation, we find that it is not possible to obtain the
critical horizon radius, $r_{+c}$, analytically. As a result, we will not be
able to calculate, analytically, the other critical parameters as well. So,
we use the numerical analysis in order to study the van der Waals like phase
transition of the black holes. In addition, we use such numerical analysis
for investigating the effects of different parameters on the critical
quantities.
\begin{figure}[tbp]
$%
\begin{array}{ccc}
\epsfxsize=5.5cm \epsffile{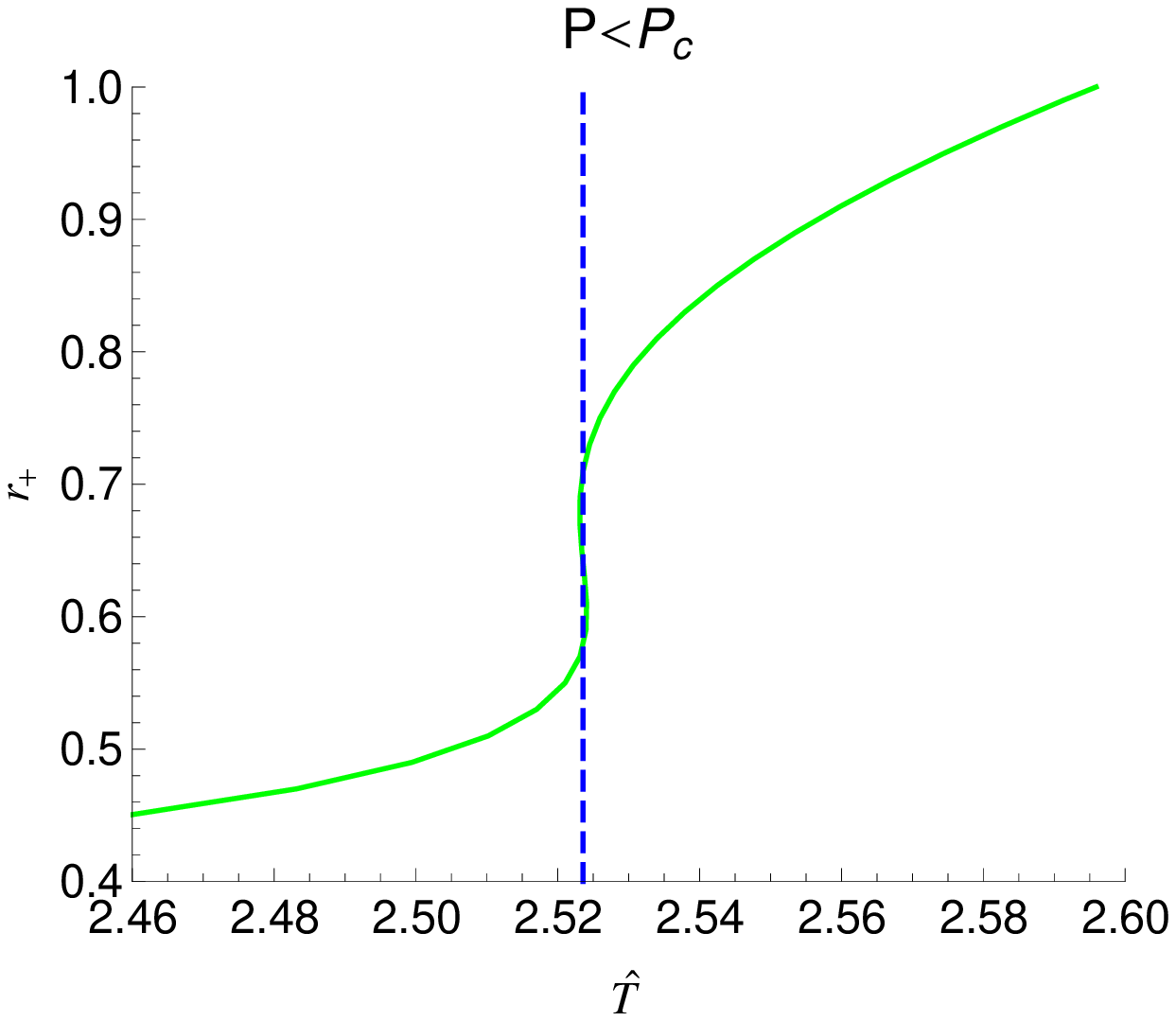} & \epsfxsize=5.5cm \epsffile{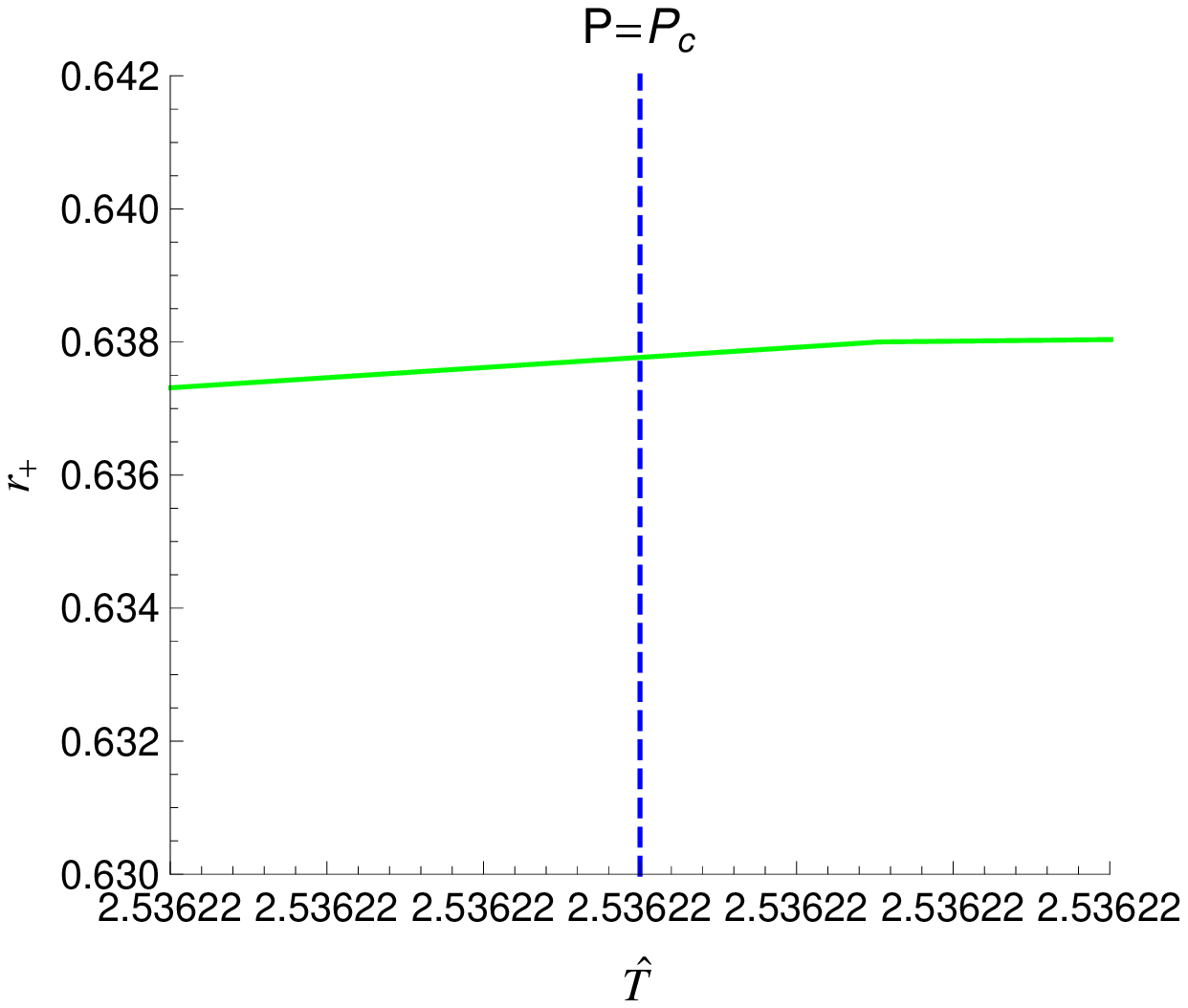}
& \epsfxsize=5.5cm \epsffile{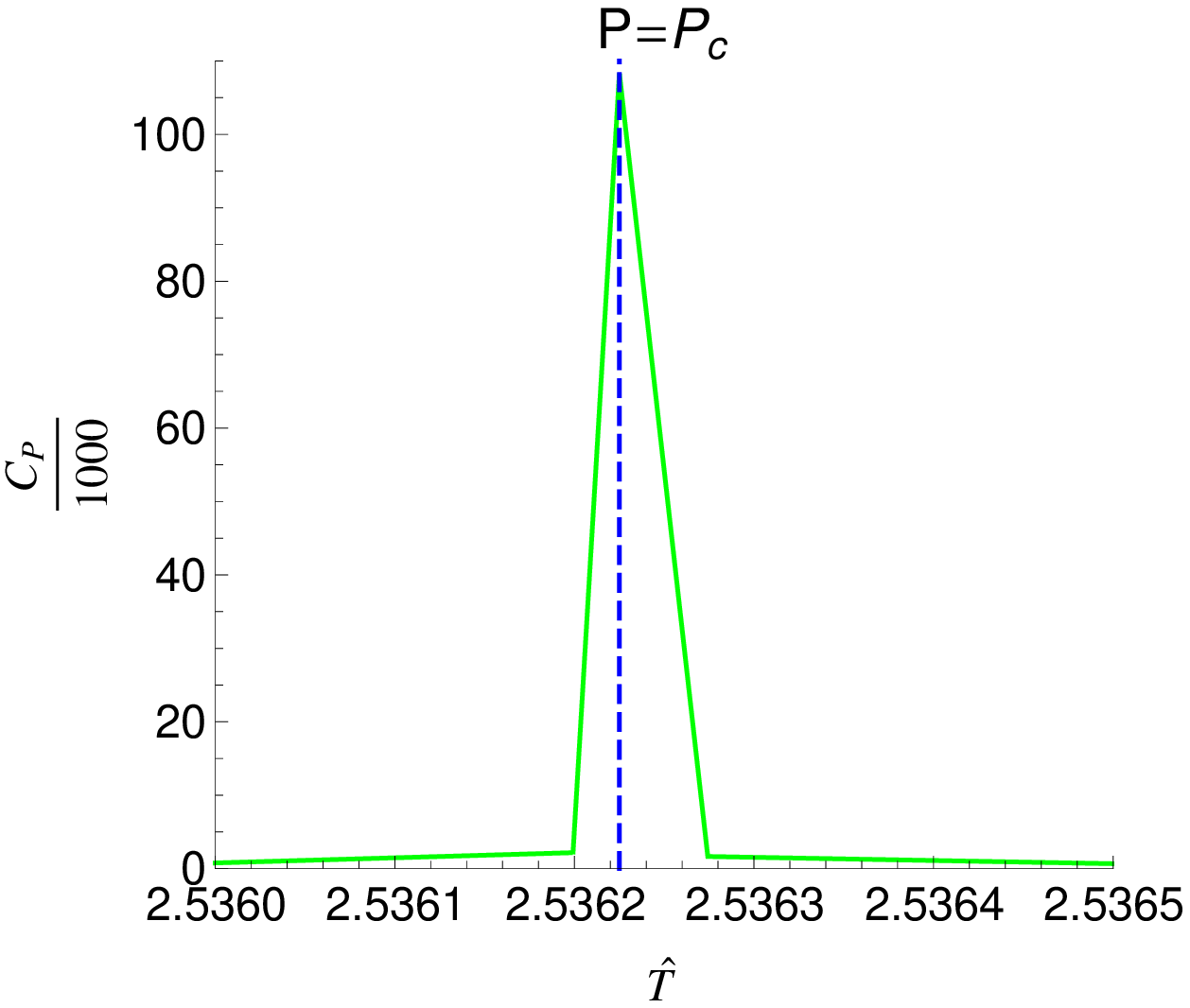}%
\end{array}
$%
\caption{$r_{+}-\hat{T}$ and $C_{P}-\hat{T}$ diagrams for $\protect\nu =1$, $%
q=2$, $\protect\beta =1$, $m=3$, $c=1$, and $c_{2}=2$. The vertical dashed
line in the left panel represents the temperature of the phase transition
point ($0.995\hat{T}_{c}$), and in the middle and right panels indicates the
critical temperature, $\hat{T}_{c}$. The discontinuity is present in the
first differential of the Gibbs free energy at phase transition point in the
left panel (due to existence of latent heat) which shows SBH and LBH undergo
a first order phase transition for $P<P_{c}$. Continuous behavior of volume
versus temperature (middle panel) and the existence of a sharp spike (weak
singularity) in the specific heat at $\hat{T}_{c}$ indicate that the system
enjoys a second order phase transition at critical point.}
\label{FirstOrder}
\end{figure}
\begin{figure}[tbp]
$%
\begin{array}{ccc}
\epsfxsize=5.5cm \epsffile{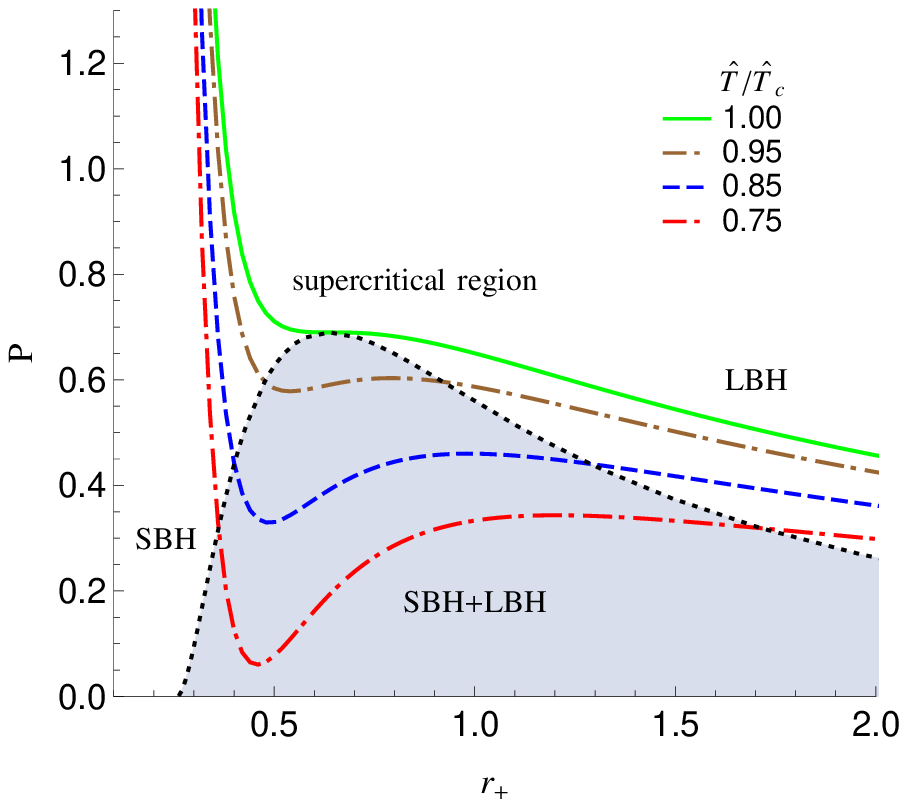} & \epsfxsize=5.5cm \epsffile{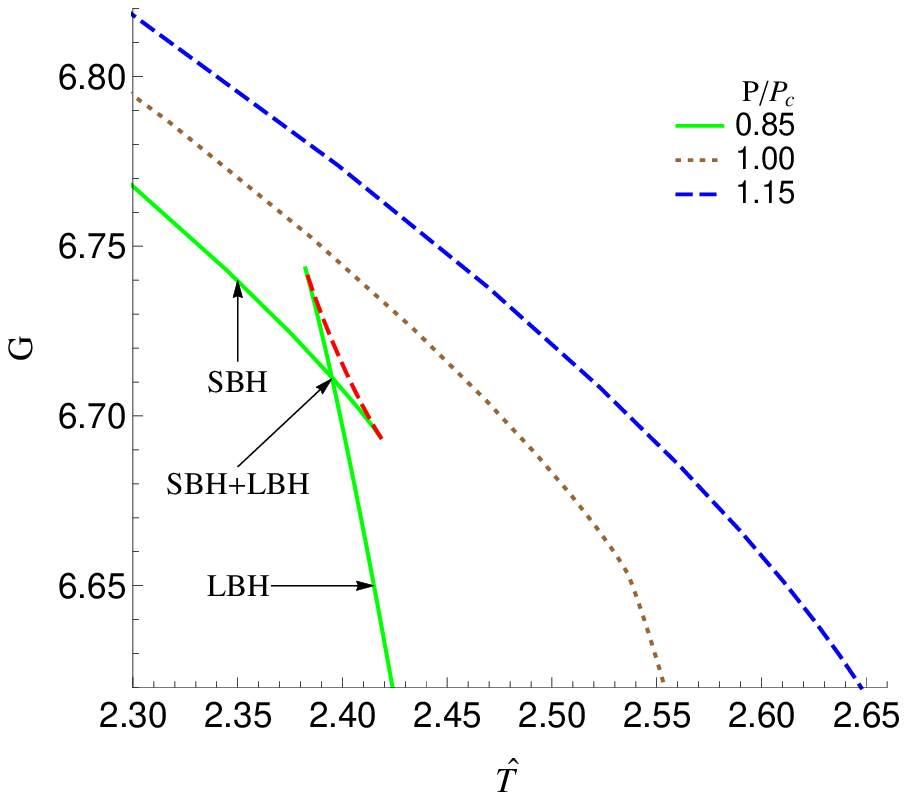} & %
\epsfxsize=5.5cm \epsffile{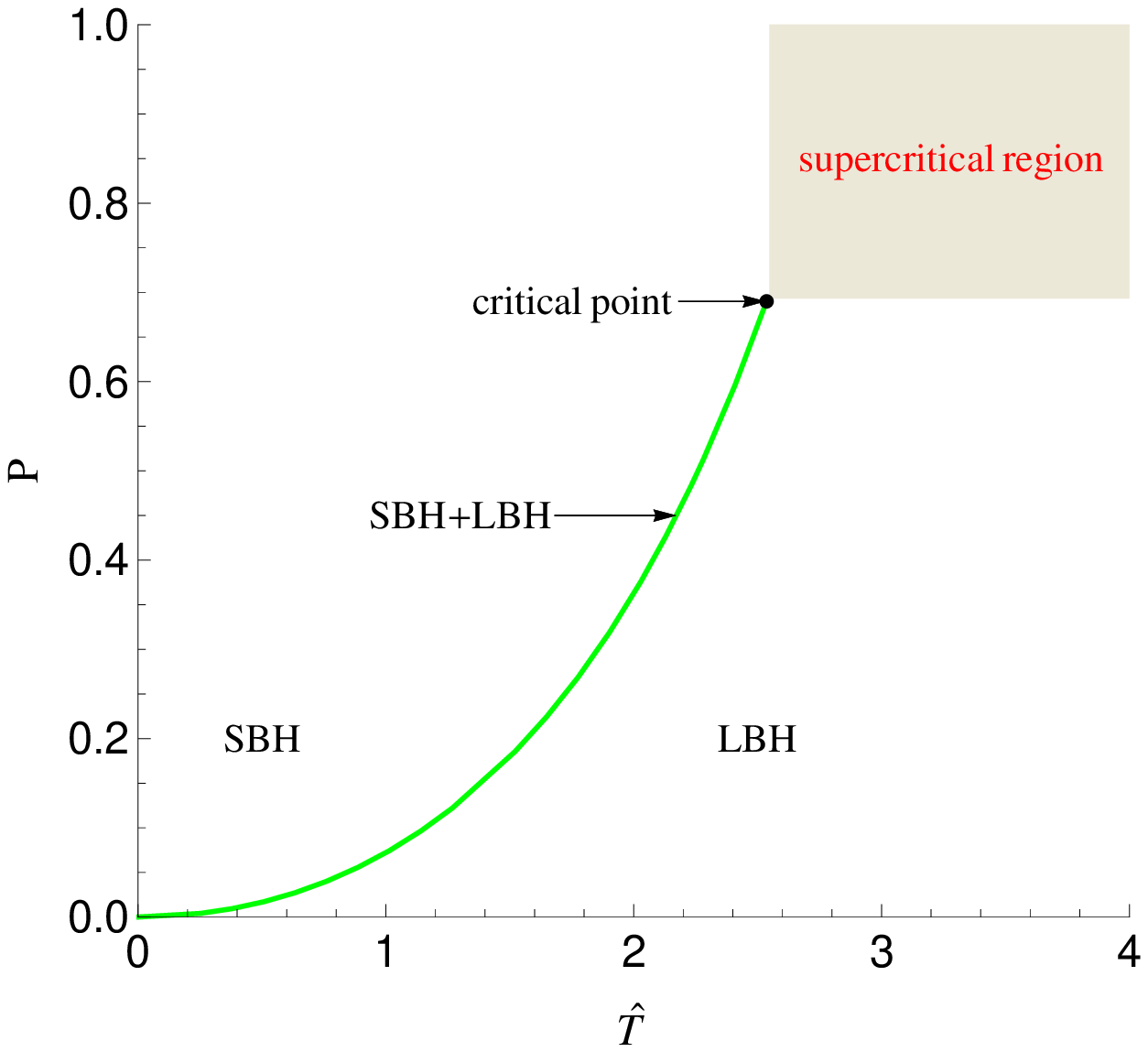}%
\end{array}
$%
\caption{$P-r_{+}$, $G-\hat{T}$, and $P-\hat{T}$ diagrams for $\protect\nu %
=1 $, $q=2$, $\protect\beta =1$, $c=1$, $c_{2}=2$, and $m=3$.}
\label{PVfig}
\end{figure}
\begin{figure}[tbp]
$%
\begin{array}{ccc}
\epsfxsize=7.5cm \epsffile{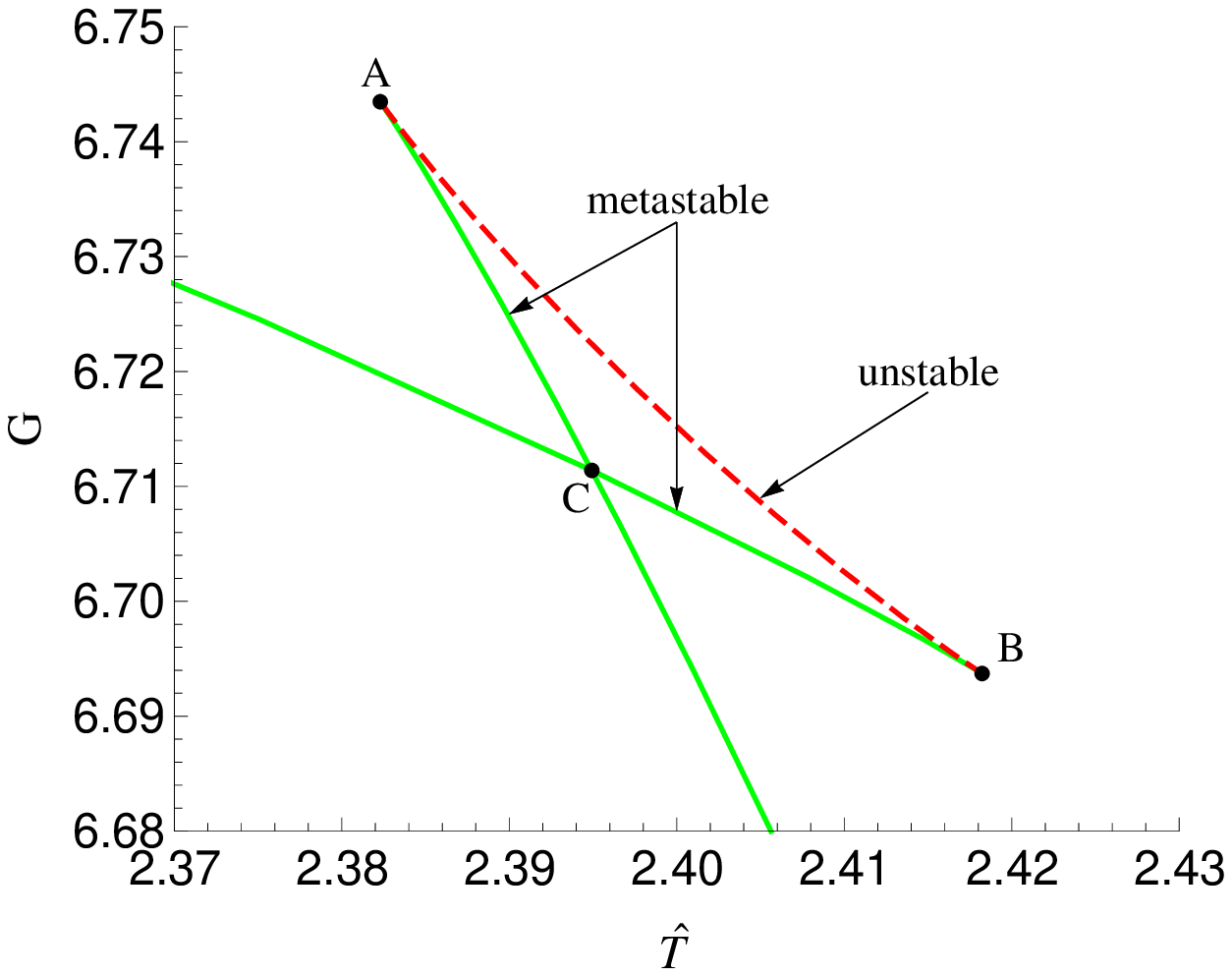} & \epsfxsize=7.5cm \epsffile{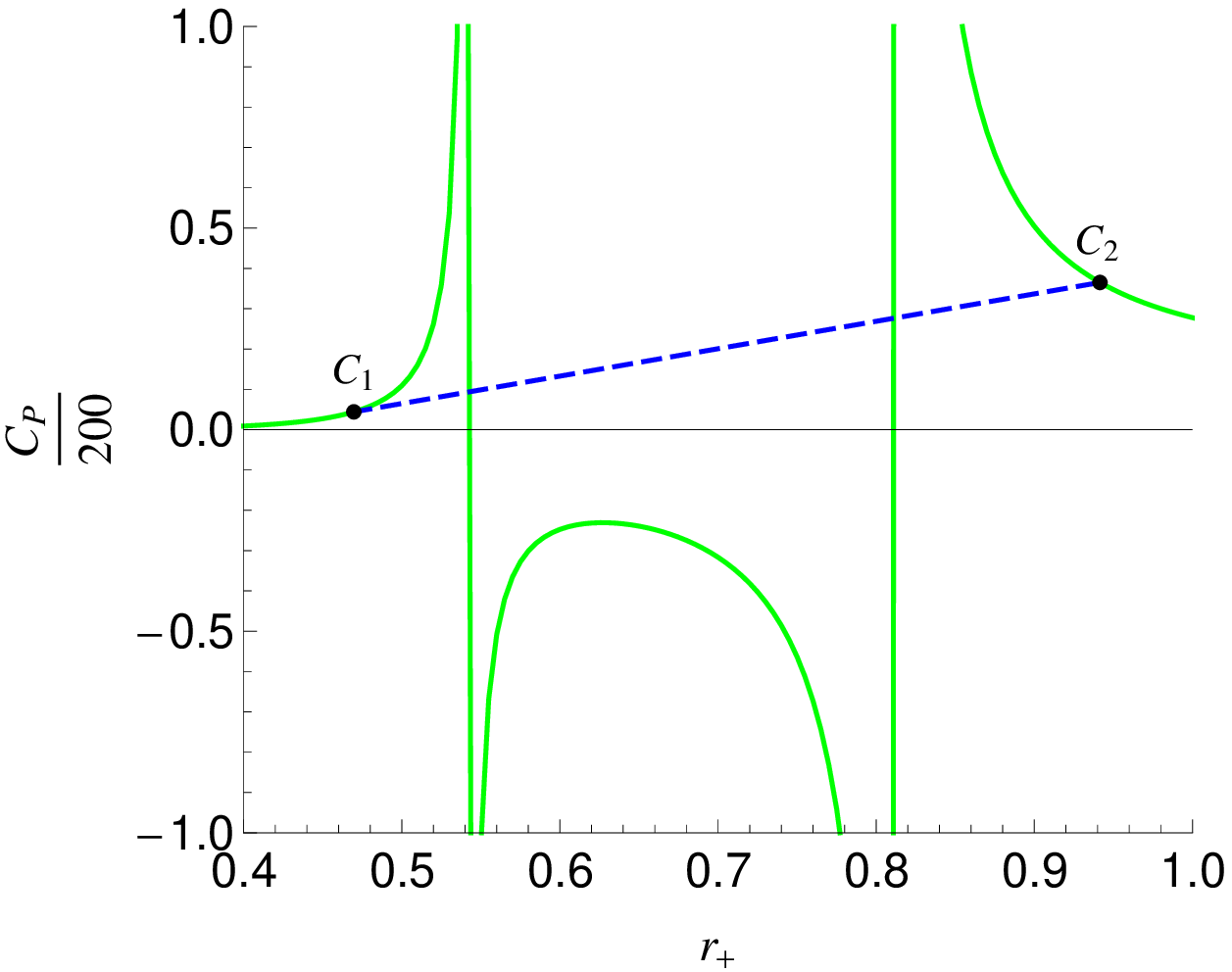} &
\end{array}
$%
\caption{ $G-\hat{T}$ and $C_{P}-r_{+}$ diagrams for $P=0.85P_{c}$, $\protect%
\nu =1$, $q=2$, $\protect\beta =1$, $c=1$, $c_{2}=2$, and $m=3$. The path $%
A-B$ indicates unstable black holes which is equivalence to the negative
heat capacity between two divergencies. The path $A-C$ ($B-C$) indicates
metastable black holes which is equivalence to the positive heat capacity
between the larger (smaller) divergency and $C_{2}$ ($C_{1}$). The SBH-LBH
phase transition occurs at point $C$ in $G-\hat{T}$ diagram, and a jump
between points $C_{1}$ and $C_{2}$ in $C_{P}-r_{+}$ diagram.}
\label{Compare}
\end{figure}
\begin{figure}[tbp]
$%
\begin{array}{ccc}
\epsfxsize=8cm \epsffile{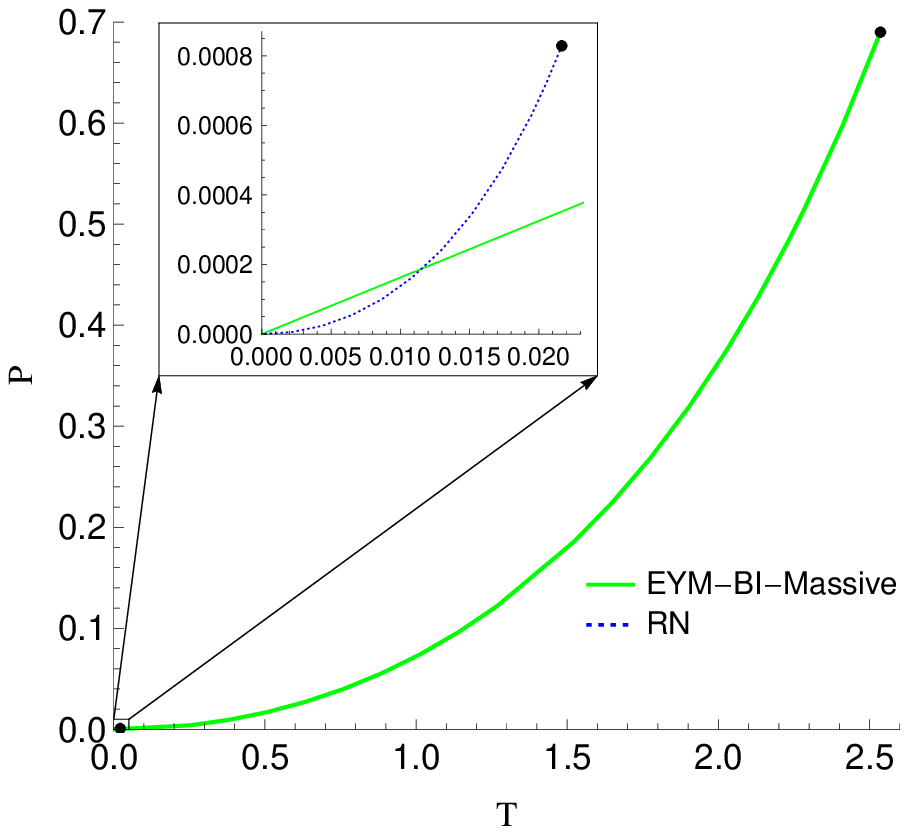} &  &
\end{array}
$%
\caption{The coexistence curve of EYM-BI-Massive and Reissner-Nordstr\"{o}m
black holes for $\protect\nu =1$, $q=2$, $\protect\beta =1$, $c=1$, $c_{1}=0$%
, $c_{2}=2$, and $m=3$.}
\label{PTRN}
\end{figure}

Paul Ehrenfest has categorized the phase transition of thermodynamical
systems based on the discontinuity in derivatives of the Gibbs free energy.
The order of a phase transition is the order of the lowest differential of
the Gibbs free energy that shows a discontinuity at the phase transition
point. Thus, in a first order phase transition, there exists a discontinuity
in the first derivative of $G$ (the entropy or volume). Next, in a second
order phase transition, the entropy or volume becomes a continuous function
and the heat capacity which is given by
\begin{equation}
C_{P}=\hat{T}\left( \frac{\partial S\ }{\partial \hat{T}}\right) _{P}=\frac{8%
\hat{T}\pi ^{2}r_{+}^{5}\sqrt{q^{2}+\beta ^{2}r_{+}^{4}}}{2\beta
r_{+}^{2}\left( q^{2}-\beta ^{2}r_{+}^{4}\right) +\sqrt{q^{2}+\beta
^{2}r_{+}^{4}}\left[ 3\nu ^{2}+2r_{+}^{4}\left( 4\pi P+\beta ^{2}\right)
-r_{+}^{2}\left( 1+m^{2}c^{2}c_{2}\right) \right] }  \label{CP}
\end{equation}%
shows a sharp spike. Clearly, Fig. \ref{FirstOrder} confirms that the black
holes under consideration enjoy the first order phase transition for
temperatures and pressures less than their critical values and they undergo
a second order phase transition at the critical point.

For instance, we plot $P-r_{+}$ isotherm, $G-\hat{T}$, and $P-\hat{T}$
diagrams for some fixed parameters to show the general phase transition
behavior of the solutions (Fig. \ref{PVfig}). Considering Fig. \ref{PVfig},
we find that the obtained black holes have a van der Waals like phase
transition between small black holes (SBH) and large black holes (LBH), and
therefore, they enjoy a first order SBH-LBH phase transition. In this
figure, $P-r_{+}$ isotherms show SBH area on the left, SBH+LBH coexistence
area in the middle, and LBH area on the right. The dotted curve is a
boundary between the regions of SBH, SBH+LBH, and LBH in the $P-r_{+}$
diagram. For temperatures above the critical temperature, there is no
physical distinction between SBH and LBH phases, and this area is denoted as
the supercritical region. In addition, in the $G-\hat{T}$ diagram, the phase
transition point is located at the cross point, where SBH+LBH are presented,
and black holes always choose the lowest energy. Moreover, the $P-\hat{T}$
diagram indicates the coexistence line between SBH and LBH which terminates
at the critical point. The critical point is located at the topmost of the
coexistence line with $P=P_{c}$, $r_{+}=r_{+c}$, and $\hat{T}=\hat{T}_{c}$.
If black hole crosses the coexistence line from left to right or top to
bottom, the system goes under a first order phase transition from SBH to
LBH. Above the critical point, SBH and LBH are physically indistinguishable
which is denoted by supercritical region.

From the left panel of Fig. \ref{Compare}, one can see that the red dashed
(solid green) line corresponds to the negative (positive) heat capacity at
constant pressure, $C_{P}$, in the right panel. In addition, the
divergencies of $C_{P}$ is indicated by two small black points $A$ and $B$
in the $G-\hat{T}$ diagram. The path bounded by these points is
unconditionally unstable, but the paths $A-C$ and $B-C$ are metastable.
Equivalently, in $C_{P}$ diagram, the region between point $C_{1}$ ($C_{2}$)
and smaller (larger) divergency is metastable, and SBH-LBH phase transition
does occur between $C_{1}$ and $C_{2}$. This figure shows that during the
phase transition from SBH to LBH, the heat capacity of the system increases.
Moreover, this figure confirms that in order to have SBH-LBH phase
transition, a local instability in the heat capacity is required.\

In addition, Fig. \ref{PTRN} shows that the generalization of
Einstein-Maxwell black holes into massive gravity and YM theory has a
significant effect on the Reissner-Nordstr\"{o}m black holes. In this
theory, the region of SBH and LBH increases, and therefore, there is van der
Waals like phase transition for higher temperatures and pressures compared
with Reissner-Nordstr\"{o}m black holes.

\begin{center}
\begin{tabular}{ccccccccccccc}
\hline\hline
$q$ &  & $\beta $ &  & $\nu $ &  & $m$ &  & $r_{+c}$ &  & $\hat{T}_{c}$ &  &
$P_{c}$ \\ \hline\hline
$2.0$ &  & $1.0$ &  & $1.0$ &  & $3.0$ &  & $0.6374$ &  & $2.5362$ &  & $%
0.6900$ \\ \hline
$2.1$ &  & $1.0$ &  & $1.0$ &  & $3.0$ &  & $0.6416$ &  & $2.4855$ &  & $%
0.6694$ \\ \hline
$2.2$ &  & $1.0$ &  & $1.0$ &  & $3.0$ &  & $0.6460$ &  & $2.4353$ &  & $%
0.6492$ \\ \hline
$2.0$ &  & $1.2$ &  & $1.0$ &  & $3.0$ &  & $0.6603$ &  & $2.3547$ &  & $%
0.5966$ \\ \hline
$2.0$ &  & $1.4$ &  & $1.0$ &  & $3.0$ &  & $0.6893$ &  & $2.1881$ &  & $%
0.5106$ \\ \hline
$2.0$ &  & $1.0$ &  & $1.1$ &  & $3.0$ &  & $0.7035$ &  & $2.3138$ &  & $%
0.5652$ \\ \hline
$2.0$ &  & $1.0$ &  & $1.2$ &  & $3.0$ &  & $0.7708$ &  & $2.1305$ &  & $%
0.4718$ \\ \hline
$2.0$ &  & $1.0$ &  & $1.0$ &  & $3.1$ &  & $0.6120$ &  & $2.8472$ &  & $%
0.8145$ \\ \hline
$2.0$ &  & $1.0$ &  & $1.0$ &  & $3.2$ &  & $0.5888$ &  & $3.1814$ &  & $%
0.9538$ \\ \hline
\end{tabular}%
\\[0pt]

Table $IV$: The effects of different parameters on the critical values of
the horizon radius, temperature, and pressure for $c=1$ and $c_{2}=2$.
\end{center}

In order to study the effects of different parameters on the critical
points, we take table $IV$ based on the numerical analysis. It is worthwhile
to mention that by increasing the critical temperature and pressure, the
region of SBH and LBH increases, and therefore, the region of phase
transition increases too. From table $IV$, we find that the critical horizon
radius is a decreasing function of the massive parameter and the critical
temperature and pressure are increasing functions of this parameter.
Considering table $IV$, one can see opposite behavior for the other
parameters such as $q$, $\beta $, and $\nu $. In other words, the critical
horizon radius is an increasing function of these parameters, whereas the
critical temperature and pressure are decreasing functions of them.

\section{adS/CFT correspondence}

In this section, we are going to point out two applications of the obtained
solutions in the context of the adS/CFT correspondence. The adS/CFT
correspondence relates string theory on asymptotically adS spacetimes to a
conformal field theory on the boundary \cite{Maldacena}. It is well-known
that this holographic correspondence between a quantum field theory and a
gravitational theory can be extended to explain some aspects of nuclear
physics \cite{Mateos}. In addition, some phenomena like the Nernst effect
\cite{Kovtun,Herzog}, superconductivity \cite{Horowitz}, Hall effect \cite%
{Hartnoll} and the decaying time scale of perturbations of a thermal state
in the field theory \cite{Hubeny} have dual descriptions in gravitational
theory.

\subsection{Holographic superconductors}

Here, we give some tips regarding the holographically dual superconductors
of the Lagrangian (\ref{Action}). First of all, one should note that at the
boundary ($r\rightarrow \infty $), the metric function (\ref{metric function}%
)\ tends to
\begin{equation}
f(r)=1-\frac{m_{0}}{r}-\frac{\Lambda r^{2}}{3}+\frac{\nu ^{2}}{r^{2}}+\frac{%
m^{2}}{2r}\left( cc_{1}r^{2}+2c^{2}c_{2}r\right) +\frac{q^{2}}{r^{2}}+%
\mathcal{O}\left( \frac{1}{r^{6}}\right) ,
\end{equation}%
and we find that the nonlinearity parameter $\beta $ does not play a
significant role in the conductivity. Therefore, the BI NED can be replaced
by Maxwell electrodynamics and the proper Lagrangian takes the following
form
\begin{equation}
\mathcal{L}=\mathcal{R}-2\Lambda -\mathcal{F}_{M}-\mathcal{F}%
_{YM}+m^{2}\sum_{i}c_{i}\mathcal{U}_{i}(g,f).
\end{equation}

In this case, due to the presence of Maxwell and YM fields, there are two
options to investigate the holographic superconductors based on perturbing
either Maxwell field or YM field. If we perturb the Maxwell (YM) field, the
YM (Maxwell) field can be considered as an extra filed that is added to the
Lagrangian as a matter source. If one wants to choose the Maxwell field to
investigate the holographic superconductors, the case will be very similar
to \cite{Vegh} (except the extra YM field they are the same) and it can be
followed. Otherwise, if the YM field is preferred to describe the
conductivity, the $SU(2)$\ gauge group should break down to the gauge
symmetry $U(1)_{3}$\ generated by the third\ component $t_{3}$\ of the gauge
field $SU(2)$\ \cite{Pufu} (see also \cite{Gubser,Alishahiha}). Thus, the
electromagnetic $U(1)$\ gauge symmetry is identified with the abelian $%
U(1)_{3}$\ subgroup of the $SU(2)$\ group. Therefore, $U(1)_{3}$\ is
interpreted as the gauge group of electromagnetism which is considered in
the boundary theory and Maxwell electrodynamics is an extra field.

\subsection{Quasinormal modes}

In terms of the adS/CFT correspondence, a large black hole in adS spacetime
corresponds to an approximately thermal state in conformal field theory.
Scalar perturbations of the black hole correspond to perturbations of this
state. Thus, the decay of the scalar field describes the decay of
perturbations of this thermal state. Therefore, we can calculate the time
scale for the approach to thermal equilibrium by calculating the quasinormal
modes (QNMs) of a large static black hole in asymptotically adS spacetime.
Here, we shall obtain the QNMs of constructed black hole solutions to find
the stability time scale of the corresponding thermal state. The other
advantage of calculating the QNMs is investigating the dynamical stability
of obtained black hole solutions undergoing scalar perturbations.

In order to calculate the QNMs, one can follow either Horowitz-Hubeny
approach \cite{Hubeny} or pseudospectral method \cite{Boyd}. The first one
is based on Fr\"{o}benius expansion of the modes near the event horizon and
forcing the differential equation to obey the boundary condition at the
horizon. The second method replaces the continuous variable by a discrete
set of points and solves the resulting generalized eigenvalue equation.
However, we follow the pseudospectral method and use a public code presented
in \cite{Jansen}\ to calculate the QN modes.

We now consider the fluctuations of a massless scalar field in the
background spacetime of obtained black holes. In order to use the
pseudospectral method, it is convenient to obtain the master equation in
Eddington-Finkelstein coordinates. In these coordinates, the background line
element takes the form%
\begin{eqnarray}
ds^{2} &=&-f(u)dt^{2}-2u^{-2}dtdu+u^{-2}\left( d\theta ^{2}+\sin ^{2}\theta
d\varphi ^{2}\right) , \\
f(u) &=&1-2Mu+\frac{1}{u^{2}L^{2}}+\nu ^{2}u^{2}+\frac{m^{2}}{2u}\left(
cc_{1}+2c^{2}c_{2}u\right) +\frac{2\beta ^{2}}{3u^{2}}\left( 1-\mathcal{%
\tilde{H}}_{1}\right) ,  \label{NMF}
\end{eqnarray}%
where $\tilde{H}_{1}={}_{2}F_{1}\left( -\frac{1}{2},-\frac{3}{4};\frac{1}{4}%
;-\frac{q^{2}u^{4}}{\beta ^{2}}\right) $, $L$\ is the adS radius related to
the cosmological constant by $\Lambda =-3/L^{2}$, and $u=1/r$. Thus, $u=0$\
corresponds to the boundary and $u=1$\ represents the horizon. The equation
of motion for a minimally coupled scalar field is governed by the
Klein-Gordon equation%
\begin{equation}
\square \Phi =0.  \label{SP}
\end{equation}

It is convenient to expand the scalar field eigenfunction $\Phi $\ in the
form
\begin{equation}
\Phi \left( t,u,\theta ,\varphi \right) =\sum_{\ell m}\psi \left( u\right)
Y_{\ell m}\left( \theta ,\varphi \right) e^{-i\omega t},  \label{EXP}
\end{equation}%
where $Y_{lm}\left( \theta ,\varphi \right) $\ denotes the spherical
harmonics. Substituting the scalar field decomposition (\ref{EXP}) into (\ref%
{SP}) leads to the following second-order differential equation for the
radial part%
\begin{equation}
u^{3}f(u)\psi ^{\prime \prime }\left( u\right) +\left[ 2i\omega
u+u^{3}f^{\prime }\left( u\right) \right] \psi ^{\prime }\left( u\right) -%
\left[ 2i\omega +u\ell \left( \ell +1\right) \right] \psi \left( u\right) =0
\label{DEq}
\end{equation}%
in which $\ell $\ is the multipole number and $\omega =\omega _{r}-i\omega
_{i}$\ is the QN frequency with an imaginary part $\omega _{i}$\ giving
damping of perturbations and a real part $\omega _{r}$\ giving oscillations.
Therefore, in terms of the adS/CFT correspondence, $\tau =1/\omega _{i}$\ is
the time scale that the thermal state needs to pass to meet the thermal
equilibrium. On the other hand, the negativity of the imaginary part
guarantees the dynamical stability of the black hole \cite{Hubeny}.
Otherwise, the perturbations increase in time and the spacetime becomes
unstable.

Causality requires ingoing modes at the event horizon and finite modes at
spatial infinity that results in a discrete spectrum of frequencies $\omega $%
. In order to analyze the behavior of modes $\psi \left( u\right) $\ near
the horizon and the spacial infinity, we set $r_{+}=1$\ and replace the
value of $M$\ by considering $f\left( r_{+}\right) =0$\ without loss of
generality. Starting with the horizon, by substituting an ansatz $\psi
\left( u\right) =\left( 1-u\right) ^{p}$\ in (\ref{DEq}), we find two
solutions as $\psi _{in}\left( u\right) \propto Const$\ and $\psi
_{out}\left( u\right) \propto \left( 1-u\right) ^{i\Omega }$\ where $\Omega
=\omega /\left( 2\pi T\right) $. By considering the time dependence $%
e^{-i\omega t}$, the $\psi _{out}\left( u\right) $\ behaves as%
\begin{equation}
\psi _{out}\left( u\right) \propto e^{-i\Omega \left[ 2\pi Tt-\ln \left(
1-u\right) \right] }.
\end{equation}

In order to keep a constant phase, $1-u$\ has to increase as $t$\ increases,
and thus $u$\ should decrease which means that this solution is outgoing.
Therefore, we must consider just the ingoing solution $\psi _{in}\left(
u\right) \propto Const$. There are two solutions near the event horizon; a
normalizable mode $\psi \left( u\right) \propto u^{3}$\ and a
non-normalizable one $\psi \left( u\right) \propto Const$. If we rescale $%
\psi \left( u\right) =u^{2}\tilde{\psi}\left( u\right) $, then the
normalizable mode tends to zero linearly, whereas the non-normalizable mode
diverges as $\sim u^{-2}$. Doing this redefinition, the wave equation (\ref%
{DEq}) becomes%
\begin{eqnarray}
&&\frac{u^{3}}{6}\left[ -6\left( \frac{1}{L^{2}}+u^{2}\left( 1+u^{2}\nu
^{2}\right) \right) -3cm^{2}u\left( c_{1}+2cc_{2}u\right) +4\beta ^{2}\left(
\mathcal{\tilde{H}}_{1}-1\right) +\mathcal{A}\right] \tilde{\psi}^{\prime
\prime }\left( u\right)   \notag \\
&&+\left[ -\frac{10}{3u_{+}^{3}}\beta ^{2}u^{5}\mathcal{\tilde{H}}_{1+}+%
\frac{u^{2}}{6}\left( \frac{6\left( 5u^{3}-2\right) }{L^{2}}+20\beta ^{2}%
\mathcal{\tilde{H}}_{1}+\mathcal{B}+3u\mathcal{C}\right) \right] \tilde{\psi}%
^{\prime }\left( u\right)   \notag \\
&&+\left[ \frac{4}{3}\beta ^{2}u\left( 1+2u_{+}^{-3}u^{3}-3\sqrt{1+\frac{%
q^{2}u^{4}}{\beta ^{2}}}\right) +\frac{8}{3}\beta ^{2}u\left( \mathcal{%
\tilde{H}}_{1}-u_{+}^{-3}u^{3}\mathcal{\tilde{H}}_{1+}\right) +u\mathcal{D}%
\right] \tilde{\psi}\left( u\right)
\begin{array}{c}
=0%
\end{array}%
,  \label{deq}
\end{eqnarray}%
where%
\begin{eqnarray}
\mathcal{A} &=&u_{+}u^{3}\left\{ 6\nu ^{2}-4\beta ^{2}u_{+}^{-4}\mathcal{%
\tilde{H}}_{1+}+u_{+}^{-2}\left[ 6+2u_{+}^{-2}\left( 2\beta ^{2}+\frac{3}{%
L^{2}}\right) +3cm^{2}\left( c_{1}u_{+}^{-1}+2cc_{2}\right) \right] \right\}
, \\
\mathcal{B} &=&-4\beta ^{2}\left( 2-5u_{+}^{-3}u^{3}+3\sqrt{1+\frac{%
q^{2}u^{4}}{\beta ^{2}}}\right) , \\
\mathcal{C} &=&-4i\Omega +cm^{2}\left[ 2cc_{2}u\left( 5u_{+}^{-1}u-4\right)
+c_{1}\left( 5u_{+}^{-2}u^{2}-3\right) \right] +2u\left[ 5u_{+}^{-1}u-4+\nu
^{2}uu_{+}\left( 5-6u_{+}^{-1}u\right) \right] , \\
\mathcal{D} &=&2\nu ^{2}u^{3}u_{+}\left( 2-\frac{3u}{u_{+}}\right) +\frac{%
2+4u_{+}^{-3}u^{3}}{L^{2}}-2i\Omega u+u^{2}\left\{ \ell \left( \ell
+1\right) -2+\frac{4u}{u_{+}}+2cm^{2}\left[ \frac{c_{1}u}{u_{+}^{2}}%
+cc_{2}\left( \frac{2u}{u_{+}}-1\right) \right] \right\} .
\end{eqnarray}

Now, the normalizable mode behaves smoothly at the boundary and it should be
considered, while we discard the other solution. The wave equation (\ref{deq}%
) is an input for the code and one can fix the free parameters and the event
horizon radius $r_{+}=u_{+}^{-1}$\ to calculate the QN modes.

In the previous section, it was shown that the small black holes are
unstable and non-physical, whereas the large black holes are physical and
enjoy thermal stability (see Eqs. (\ref{H1}) and (\ref{H2})). On the other
hand, the large black holes correspond to the thermal states in CFT. Thus,
we shall focus on the QNMs of large black holes ($r_{+}>>L$) and discard the
small ones ($r_{+}<<L$) for $L=1$\ as the adS radius.

Here, we set the free parameters as $q=1$, $c=1$, $c_{1}=-1$, $c_{2}=2$,\
and $\ell =0$, and evaluate the QNMs for different values of $m$, $\beta $, $%
\nu $, and $r_{+}$. In table $V$, we list the QNM frequencies\ for the
fundamental\ mode ($n=0$) and the first overtone ($n=1$) of intermediate
black holes ($r_{+}=5,10$) and large ones ($r_{+}=50,100$). From this table,
one can see that as the overtone number and the event horizon radius
increase, both the real and imaginary parts of frequencies increase as well.
But an opposite behavior is seen for increasing in the graviton mass.
Besides, the real (imaginary) part of the frequencies decreases (increases)
when the magnetic charge increases. We recall that the nonlinearity
parameter $\beta $\ does not play a significant role at the boundary $%
r\rightarrow \infty $\ ($u\rightarrow 0$), and thus increasing/decreasing in
$\beta $\ does not change the value of QNMs as it can be seen from the
table. Therefore, the BI NED can be replaced by Maxwell electrodynamics when
we want to investigate the applications of the solutions in the context of
the adS/CFT correspondence. We should mention that as the imaginary part of
frequencies increases, the corresponds thermal state meets the stability
faster. In addition, the obtained black hole solutions undergoing massless
scalar perturbations are dynamically stable since all the frequencies have a
negative imaginary part.

\begin{center}
\begin{tabular}{ccccccccccccc}
\hline\hline
$m$ &  & $\beta $ &  & $\nu $ &  & $r_{+}=5$ &  & $r_{+}=10$ &  & $r_{+}=50$
&  & $r_{+}=100$ \\ \hline\hline
$2$ &  & $1$ &  & $1$ &  & $%
\begin{array}{c}
7.3348-11.5624i \\
12.7580-21.6360i%
\end{array}%
$ &  & $%
\begin{array}{c}
15.5281-24.8128i \\
26.8225-46.0131i%
\end{array}%
$ &  & $%
\begin{array}{c}
88.6935-131.3863i \\
151.9287-242.6142i%
\end{array}%
$ &  & $%
\begin{array}{c}
181.0693-264.5858i \\
309.8333-488.4349i%
\end{array}%
$ \\ \hline
$3$ &  & $1$ &  & $1$ &  & $%
\begin{array}{c}
4.8000-9.1811i \\
8.5936-17.5565i%
\end{array}%
$ &  & $%
\begin{array}{c}
11.6144-22.2860i \\
20.4356-41.6669i%
\end{array}%
$ &  & $%
\begin{array}{c}
83.8994-129.0660i \\
144.1318-238.5051i%
\end{array}%
$ &  & $%
\begin{array}{c}
176.1904-262.3050i \\
301.9004-484.3753i%
\end{array}%
$ \\ \hline
$2$ &  & $5$ &  & $1$ &  & $%
\begin{array}{c}
7.3348-11.5624i \\
12.7580-21.6361i%
\end{array}%
$ &  & $%
\begin{array}{c}
15.5281-24.8128i \\
26.8225-46.0131i%
\end{array}%
$ &  & $%
\begin{array}{c}
88.6935-131.3863i \\
151.9287-242.6142i%
\end{array}%
$ &  & $%
\begin{array}{c}
181.0693-264.5858i \\
309.8333-488.4349i%
\end{array}%
$ \\ \hline
$2$ &  & $1$ &  & $2$ &  & $%
\begin{array}{c}
7.3065-11.5752i \\
12.6996-21.6612i%
\end{array}%
$ &  & $%
\begin{array}{c}
15.5244-24.8146i \\
26.8149-46.0167i%
\end{array}%
$ &  & $%
\begin{array}{c}
88.6935-131.3863i \\
151.9287-242.6143i%
\end{array}%
$ &  & $%
\begin{array}{c}
181.0692-264.5858i \\
309.8333-488.4349i%
\end{array}%
$ \\ \hline
\end{tabular}%
\textbf{\\[0pt]
}

Table $V$: The fundamental\ mode (first line) and the first overtone (second
line) of the QN frequencies for different values of $m$, $\beta $, $\nu $,
and $r_{+}$.
\end{center}

\begin{figure}[tbp]
$%
\begin{array}{ccc}
\epsfxsize=7.5cm \epsffile{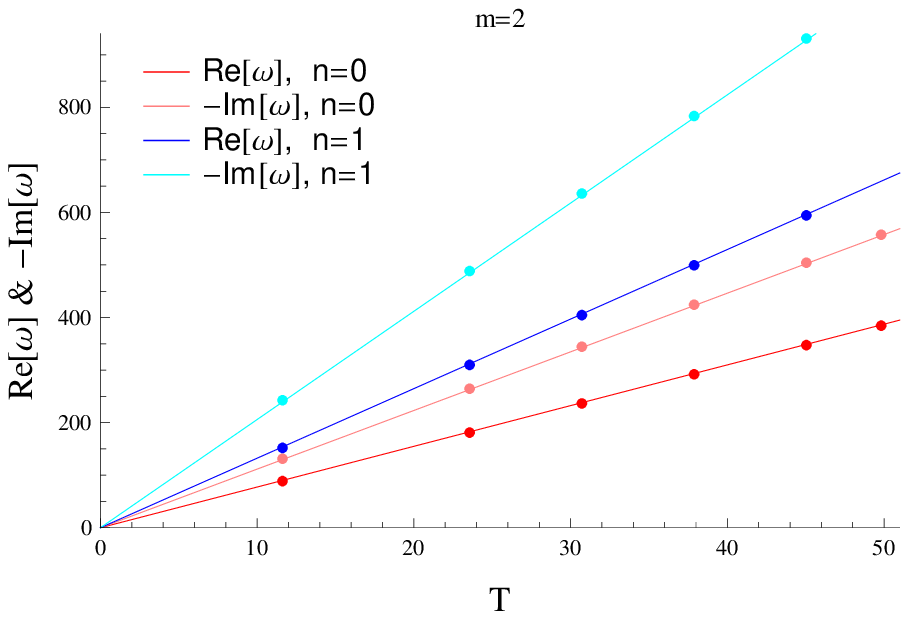} & \epsfxsize=7.5cm \epsffile{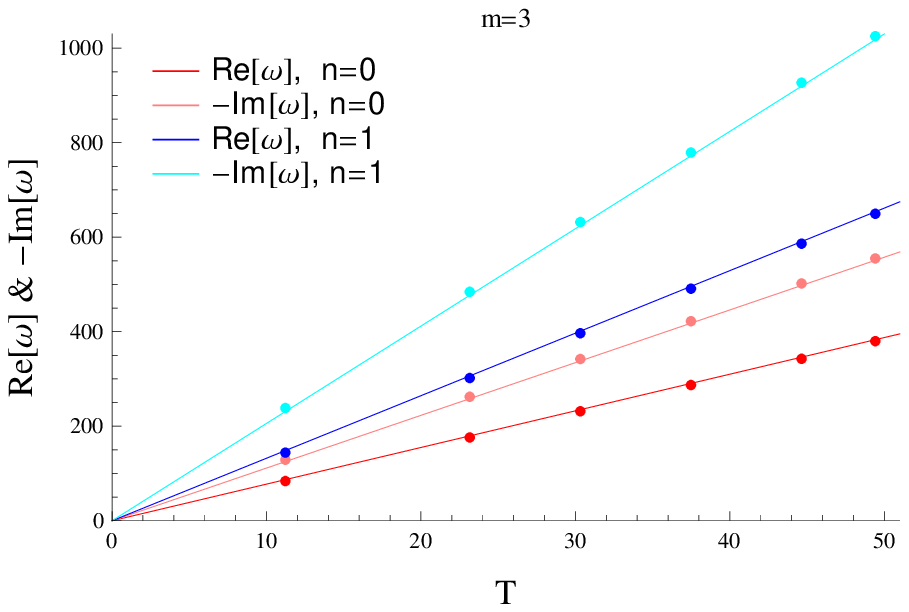} &
\end{array}
$%
\caption{ The QN frequencies for the fundamental mode and the first overtone
for $m=2$ (left panel), $m=3$ (right panel), $q=2$, $\protect\beta =1$, $%
\protect\nu =1$, $\ell =0$, $c=1$, $c_{1}=-1$, and $c_{2}=2$.}
\label{QNMs}
\end{figure}

It is worthwhile to mention that as $r_{+}$\ increases, changing in $\nu $\
does not affect the QNMs significantly (compare the first line and last line
for $r_{+}=50,100$\ in table $V$). But this is not correct in the case of $m$%
\ (compare the first line and second line for $r_{+}=50,100$). In order to
explain this fact, one may consider the temperature (\ref{T+}) for large
black holes at the first step%
\begin{equation}
T=\frac{3r_{+}+cc_{1}m^{2}}{4\pi }+\mathcal{O}\left( \frac{1}{r_{+}}\right) ,
\label{qnmT}
\end{equation}%
and secondly, look at the relation between the QNMs and this temperature
illustrated in Fig. \ref{QNMs}. As one can see, both the real and imaginary
parts of frequencies increase linearly with increase in the temperature (\ref%
{qnmT}). Therefore, changing in $\nu $\ does not affect the QNMs since it is
absent in (\ref{qnmT}), whereas $m$\ is present. From (\ref{qnmT}), we can
find that increasing in $c$\ and $c_{1}$\ leads to increasing in QNMs, but $q
$\ and $c_{2}$\ do not change the QNMs in the case of large black holes, as $%
\nu $\ did not. The points in Fig. \ref{QNMs}, representing the QNMs, lie on
straight lines through the origin. For the real part, the lines are given by%
\begin{eqnarray}
\left\{
\begin{array}{cc}
\omega _{r}=7.747T, & n=0 \\
\omega _{r}=13.236T, & n=1%
\end{array}%
\right. \text{ for }m &=&2,  \label{Rp1} \\
\left\{
\begin{array}{cc}
\omega _{r}=7.752T, & n=0 \\
\omega _{r}=13.230T, & n=1%
\end{array}%
\right. \text{ for }m &=&3,  \label{Rp2}
\end{eqnarray}%
while for the imaginary part we have%
\begin{eqnarray}
\left\{
\begin{array}{cc}
\omega _{i}=11.158T, & n=0 \\
\omega _{i}=20.594T, & n=1%
\end{array}%
\right. \text{ for }m &=&2,  \label{Ip1} \\
\left\{
\begin{array}{cc}
\omega _{i}=11.151T, & n=0 \\
\omega _{i}=20.596T, & n=1%
\end{array}%
\right. \text{ for }m &=&3.  \label{Ip2}
\end{eqnarray}

In terms of the adS/CFT correspondence, $\tau =1/\omega _{i}$\ is the time
scale for the approach to thermal equilibrium. Therefore, Eqs. (\ref{Ip1})
and (\ref{Ip2}) are the main results of this subsection. One may note that
both the real and the imaginary parts of the frequencies are linear
functions of $r_{+}$\ since the temperature of large black holes is a linear
function of $r_{+}$. Interestingly, the same result was found for the
Schwarzschild-adS black hole \cite{Hubeny}.

\section{Conclusions \label{Conclusions}}

In this paper, we have obtained Einstein-Massive black hole solutions in the
presence of YM and BI NED fields. We have also studied the geometric
properties of the solutions and it was shown that there is an essential
singularity at the origin which can be covered with an event horizon. In
addition, we have calculated the conserved and thermodynamical quantities,
and it was shown that even though the YM and BI NED fields modify the
solutions, the first law of thermodynamics is still valid.

Moreover, we have studied the thermal stability of the obtained black holes
and investigated the effects of different parameters on the stability
conditions. We have found that the large black holes ($r_{+}>r_{+\max }$)
are physical and stable, whereas the small black holes ($r_{+}<r_{+\min }$)
are non-physical ($T<0$). Furthermore, we have classified the medium black
holes ($r_{+\min }<r_{+}<r_{+\max }$) in Fig. \ref{CQadS} and investigated
the effects of different parameters on thermal stability of these black
holes in tables $I-III$.

In addition, we have considered the cosmological constant as thermodynamical
pressure and it was shown that the obtained black holes enjoy the first
order SBH-LBH phase transition. Also, we have studied this kind of phase
transition in the heat capacity diagram and specified the unstable and
metastable phases of obtained black holes related to the negative and
positive heat capacities, respectively. It was shown that during the phase
transition from SBH to LBH, the heat capacity of the system increases. We
have seen that the generalization of Reissner-Nordstr\"{o}m solutions into
massive gravity and YM theory increases the critical temperature and
pressure, and as a result, the region of SBH and LBH increases. Moreover, we
have investigated the effects of different parameters on the critical
points, and we found that the parameters $q$, $\beta $, and $\nu $ have
opposite effect on the critical points compared with the massive parameter, $%
m$.

Besides, we have considered massless scalar perturbations in the background
of obtained black holes in asymptotically adS spacetime. We also have
calculated the QN frequencies by using the pseudospectral method in order to
investigate the dynamical stability of the black holes, the effects of
different parameters on the QNMs, and obtain the time scale of the thermal
state for the approach to thermal equilibrium in CFT. It was seen that the
obtained solutions are dynamically stable and BI NED generalization does not
affect the frequencies. Furthermore, it was shown that increasing in $r_{+}$%
, $c$, $c_{1}$, and $m$\ lead to increase in both the real and imaginary
parts of the frequencies. It is worthwhile to mention that this result
depends on the sign of $c$\ and $c_{1}$\ (through the text, we considered a
negative value for $c_{1}$,\ and therefore, increasing in $m$\ has led to
decrease in the QNMs). In addition, we have found that $\nu $, $q$, and $%
c_{2}$\ do not affect the QNMs in the case of large black holes.\ Since a
static large black hole in adS spacetime corresponds to an approximately
thermal state in conformal field theory, $\nu $, $q$, and $c_{2}$\ have no
effect on the time scale of the thermal state. Just like the
Schwarzschild-adS black holes \cite{Hubeny}, both the real and imaginary
parts of frequencies for the large black holes were linear functions of the
temperature.

As a final remark, it is worth mentioning that although we consider the ADM
mass in the context of black hole thermodynamics, there is another extension
of mass (so-called hairy mass) for hairy black holes which is related to the
calculation of the null circular geodesic (photon-sphere) \cite{B1,B2}. Such
a hairy mass is not related to our discussion in this paper and it can be
considered as a new work with photon-sphere concentration.

\begin{acknowledgements}
The authors wish to thank the anonymous referee for the
constructive comments that enhanced the quality of this paper. We
wish to thank Shiraz University Research Council. This work has
been supported financially by the Research Institute for Astronomy
and Astrophysics of Maragha, Iran.
\end{acknowledgements}

\appendix

\section{EYM-Maxwell black holes in massive gravity}

Here, we give a brief study regarding the $P-V$ criticality of EYM-Maxwell
black holes in massive gravity. In order to find the related equation of
state, one can use the expansion of the metric function (\ref{metric
function}) for a large value of nonlinearity parameter, $\beta $, and follow
the same procedure given in Sec. \ref{PV}, which leads to
\begin{equation}
P(r_{+},T)=\frac{T}{2r_{+}}-\frac{1}{8\pi r_{+}^{2}}\left[ 1-\frac{q^{2}+\nu
^{2}}{r_{+}^{2}}+m^{2}\left( r_{+}cc_{1}+c^{2}c_{2}\right) \right] .
\end{equation}

Using the definition of the inflection point (\ref{inflection point}), we
can find the critical horizon radius, temperature, and pressure as follows%
\begin{equation}
r_{+c}=\sqrt{\frac{6\left( q^{2}+\nu ^{2}\right) }{1+m^{2}c^{2}c_{2}}},
\end{equation}%
\begin{equation}
T_{c}=\frac{m^{2}cc_{1}}{4\pi }+\frac{\left[ 1+m^{2}c^{2}c_{2}\left(
2+m^{2}c^{2}c_{2}\right) \right] }{3\pi \sqrt{6\left( q^{2}+\nu ^{2}\right)
\left( 1+m^{2}c^{2}c_{2}\right) }},
\end{equation}%
\begin{equation}
P_{c}=\frac{1+m^{2}c^{2}c_{2}\left( 2+m^{2}c^{2}c_{2}\right) }{96\pi \left(
q^{2}+\nu ^{2}\right) }.
\end{equation}

Considering equations mentioned above, we find that $T_{c}$\ depends on $%
c_{1}$,\ but $r_{+c}$\ and $P_{c}$\ are independent of this parameter. This
means that for the fixed values of $r_{+c}$\ and $P_{c}$, there is infinite $%
T_{c}$ for the system depending on the value of $c_{1}$! So, in order to get
rid of this situation, we define $\frac{m^{2}cc_{1}}{4\pi }$ as a background
temperature, $T_{0}$, and rescale the critical temperature into
\begin{equation}
\hat{T}_{c}=T_{c}-T_{0}=\frac{\left[ 1+m^{2}c^{2}c_{2}\left(
2+m^{2}c^{2}c_{2}\right) \right] }{3\pi \sqrt{\left( q^{2}+\nu ^{2}\right)
\left( 1+m^{2}c^{2}c_{2}\right) }},
\end{equation}%
which shows a unique critical temperature.

\end{document}